\documentclass[aps,showpacs,onecolumn,notitlepage,floatfix,amsmath,amssymb,nofootinbib]{revtex4-1}

\usepackage{amssymb}
\usepackage{amsmath}
\usepackage{epsfig}
\usepackage{graphicx}
\usepackage{subfig}
\usepackage{color}
\usepackage{latexsym}
\usepackage{bm,latexsym}
\usepackage{mathrsfs}
\usepackage{float}
\usepackage{pbox}
\usepackage{bigints} 

\def\Journal#1#2#3#4{{#1} {\bf #2}, #3 (#4)}
\def\AA{{Astron. Astrophys.}}
\def\AIPCP{{AIP Conference Proceedings}}
\def\AJ{{Astron. J.}}
\def\AP{{Ann. Phys.}}
\def\APF{{Ann. Phys.} Fr.}
\def\APJ{{Astrophys. J.}}
\def\APJS{{Astrophys. J. Suppl. Ser.}}
\def\ARAA{{Annu. Rev. Astron. Astrophys.}}
\def\CMM{{Commun. Math. Phys.}}
\def\CPC{{Comp. Phys. Com.}}
\def\CQG{{Class. Quantum Grav.}}
\def\EPJC{{Eur. Phys. J. C}}
\def\GRG{{Gen. Relativ. Gravit.}}
\def\IJMPD{{Int. Jour. Mod. Phys. D}}
\def\JCAP{{JCAP}}
\def\JCP{{Comput. Phys}}
\def\FCP{{Fund. CosmicPhys.}}
\def\JMP{{J. Math. Phys.}}
\def\LRR{{Living Rev. Rel.}} 
\def\MNRAS{{Mon. Not. Roy. Astron. Soc.}} 
\def\NAT{{Nature}}
\def\NATL{{Nature (London)}}
\def\NAR{{New Astron. Rev.}}
\def\NIM{\em Nucl. Instrum. Methods}
\def\NIMA{{\em Nucl. Instrum. Methods} A}
\def\NPA{{Nucl. Phys.} A}
\def\NPB{{\em Nucl. Phys.} B}
\def\NPBPS{Nucl. Phys. B (Proc. Suppl.)}
\def\PLB{{Phys. Lett.}  B}
\def\PLA{{Phys. Lett.}  A}
\def\PRL{Phys. Rev. Lett.}
\def\PREV{Phys. Rev.}
\def\PRD{{Phys. Rev.} D}
\def\PR{{Phys. Rep.}}
\def\PTP{{Prog. Theor. Phys.}}
\def\SCI{{Science}}
\def\RMF{{Rev. Mex. F\'\i s.}}
\def\RMP{{Rev. Mod. Phys.}}

\def\arxiv{{arXiv}}
\def\APPh{{Astroparticle Phys.}}
\def\MNRAS{{Mon. Not. R. Astron. Soc.}}
\usepackage[usenames,dvipsnames]{xcolor}
\definecolor{orange}{cmyk}{0,0.5,1,0}
\newcommand{\santiago}[1]{\textcolor{orange}{#1}}

\selectfont{}
\begin{document}

\title{ Black bounces as 
magnetically charged  phantom regular black holes in Einstein-nonlinear electrodynamics gravity coupled to a self-interacting scalar field  }

\author{Pedro Ca\~nate} 
\email{pcannate@gmail.com }

\affiliation{Departamento de F\'isica  Te\'orica, Instituto de F\'isica, Universidade do Estado do Rio de Janeiro, Rua S\~ao Francisco Xavier 524, Maracan\~a,
CEP 20550-013, Rio de Janeiro, Brazil. \\
Programa de F\'isica, Facultad de Ciencias Exactas y Naturales, Universidad Surcolombiana, Avenida Pastrana Borrero - Carrera 1, A.A. 385, Neiva, Huila, Colombia.
}

\begin{abstract}
As previously proposed in Simpson and Visser \cite{Simpson2019}, Mazza {\it et al.} \cite{rotating_Bb}, Franzin {\it et al.} \cite{Charged_rotating_Bb}, and Lobo {\it et al.} \cite{Simpson_4},
the ``black bounce'' spacetimes are an interesting type of globally regular modifications of the ordinary black holes (such as the Kerr-Newman geometry and its particular cases) which generically contain a spacetime singularity (usually of the curvature type) at their center. 
To transforms a static, spherically symmetric and asymptotically flat black hole (SSS-AF-BH) geometry regular everywhere except its 
center of symmetry
 $r=0$ (where $r$ stands for the ``areal radius''  
 of the two-dimensional spheres of symmetry) 
and with (outer) event horizon at $r=r_{h}>0$, into a black
 bounce spacetime, is to simply 
replace $r$ with $\sqrt{\rho^{2} + a^{2}}$ and $dr$ with $d\rho$, being $\rho$ a new radial coordinate, and $a$ is some real constant nonzero. As long as $r_{h}=\sqrt{\rho^{2}_{h} + a^{2}}>|a|$, the result is a globally regular (or singularity-free) black hole spacetime (called black bounce) where 
the singularity that occurs in the ordinary SSS-AF-BH geometry at 
$r = 0$ now in the transformed geometry
turns into a regular spacetime region determined by the  
two-dimensional spheres
of symmetry of radius $|a|$,  
while the areal radius $\sqrt{\rho^{2} + a^{2}}$ always remains positive for all 
$\rho\in(-\infty,\infty)$ 
and has a minimum at $\rho=0$ given by $|a|$. 
Hence, in the transformed spacetime, the  
areal radius has a minimum, decreasing before  
and increasing after this minimum (defining two SSS-AF regions that bounce).
\\
In this work we will present several black-bounces exact solutions of General Relativity. 
Among them is a novel type of black-bounce solution, which in contrast to the Simpson-Visser type \cite{Simpson2019,rotating_Bb,Charged_rotating_Bb,Simpson_4}, does not have the Ellis wormhole metric as a particular case. The source of these solutions is linear superposition of phantom scalar fields and nonlinear electromagnetic fields. 
\end{abstract}



\maketitle

\section{Introduction}
Einstein's theory of General Relativity (GR) is a metric theory of gravity, designed to describe spacetime, which  
has become the most accepted definition of classical gravity in modern physics.\footnote{Since the foundation of GR theory in 1915, it has been subjected to innumerable tests (some of which have been passed very successfully), for instance in the solar system or binary pulsars 
(see \cite{Bertotti,Williams,Will1,Will2,Psaltis} and references therein).}  
 However, many important physical solutions of the GR field equations [among them we can mention the Kerr-Newman black hole metric and its particular cases, and the Friedmann-Lema\^itre-Robertson-Walker (FLRW)
metric or Standard Cosmological Model]  
contain a gravitational singularity\footnote{Also referred to as 
physical singularity.}  where some relevant curvature invariants (either constructed from the Riemann tensor, such as $R \equiv R^{\alpha}{}_{\alpha}$, $R_{\alpha\beta}R^{\alpha\beta}$, $R_{\alpha\beta\mu\nu}R^{\alpha\beta\mu\nu}$, or 
formed by polynomial expressions in covariant derivatives of the Riemann tensor) diverge.
Concerning the physical meaning of these types of spacetimes, since the singularities occurring in the spacetime
denote a grave issue\footnote{Also known as the singularity problem in gravitation theory.} for  
GR as their presence signals a regime where predictability breaks down and the theory does not hold (see
Refs. \cite{Hawking1976,Carroll2001,Romero2013} for a review),
a number of proposals in classical and quantum contexts
have been suggested to get rid of the singularity (see Refs.    
\cite{Narlikar1985,Rendall1996,Hayward2006,Ashtekar2009,Kuntz2020} and the references therein). 

The construction of regular black holes has been studied using nonlinear electrodynamics (NLED) as source.
Specifically, motivated by the Bardeen model \cite{Bardeen} 
several regular black holes,\footnote{The regular black holes are an interesting type of singularly-free spacetimes, which plays an important role in order to study the quantum corrected black holes (see \cite{quantum_BH} for details).} in the sense that none of them have curvature singularities,\footnote{I.e., all invariants  
constructed from the Riemann curvature tensor $\boldsymbol{\mathcal{R}}=R_{\alpha\beta\mu\nu}\boldsymbol{dx^{\alpha}\!\otimes\! dx^{\beta}\!\otimes\!dx^{\mu}\!\otimes\!dx^{\nu}}$, e.g.  $R \equiv R^{\alpha}{}_{\alpha}$, $R_{\alpha\beta}R^{\alpha\beta}$, $R_{\alpha\beta\mu\nu}R^{\alpha\beta\mu\nu}$, and similar scalars formed by polynomial expressions in covariant derivatives of the curvature tensor 
are well defined everywhere.} have been constructed in the framework of Einstein's theory of general relativity coupled to nonlinear electrodynamics (GR-NLED) \cite{Ayon_Garcia,Ayon_Garcia2000}.  
Although these   
solutions obey the weak energy condition (WEC), they can avoid the singularity theorems \cite{Penrose,Penrose_GC,Wald2,Senovilla}
because they are not globally hyperbolic. 
Let us remark that since the regular black hole metrics described in \cite{Ayon_Garcia} are constructed
with pure electric charges, this requires different NLED Lagrangians in
different parts of spacetime; see \cite{BronnikovRBH2000,BronnikovRBH2001} for details. 

On the other hand, recently, the NLED 
has been gaining attention to generate 
primordial magnetic fields in the early Universe,\footnote{
Cosmological magnetic fields have become more important
since the wealth of observations of magnetic fields in the
Universe \cite{Durrer2013,Kunze2013}.} as 
a source of gravity that can drive the Universe into a state of accelerated expansion,
and also to resolve cosmological singularity problem; see \cite{Camara2004,Novello2007,Kruglov2016,Sarkar2021} for review.


As mentioned at the beginning of the paper,
the black bounce spacetimes \cite{Simpson2019,rotating_Bb,Charged_rotating_Bb,Simpson_4} represent a new alternative
to the black hole singularity problem.
To date, all the known static, spherically symmetric, and asymptotically flat black-bounce (SSS-AF-BB) metrics belong to the Simpson-Visser type, and have the form 
\begin{equation}
\boldsymbol{ds^{2}} = - n(r)\boldsymbol{dt^{2}} + n^{-1}(r)\boldsymbol{d\rho^{2}} + r^{2} \left(\boldsymbol{d\theta^{2}} + \sin^{2}\theta \boldsymbol{d\varphi^{2}}\right),    
\end{equation}
with areal radius function $r=r(\rho)$ given by $r = \sqrt{\rho^{2} + a^{2}}$ being $a=$ constant $\in\mathbb{R}$. 
Therefore, the areal radius function has a global minimum at $\rho=0$ given by $r_{min}=r(0)=|a|$, whereas $n(r)$ is a smooth function on $\rho\in(-\infty,\infty)$, such that $n(r)\rightarrow1$ as $\rho\rightarrow\pm\infty$ [this implies the existence of two asymptotically flat regions connected at $\rho=0$, that is, one for each side of the bounce; ($-$) for the side $\rho<0$; and ($+$) for the side $\rho>0$], and where the event horizons at $\rho=\pm\rho_{h}$ [two horizons of identical area radii $r_{h}$, one on each side of the bounce,  
fulfilling $n(r_{h})=0$ and such that for all $r>r_{h}$ the $n(r)$ function must be finite nonzero and positive definite] exist whenever $\rho_{h}=\sqrt{r^{2}_{h}-a^{2}}$ is finite nonzero. Thus, starting from the region $\rho>0$, since the spacetime is regular at $\rho = 0$, this can be extended to $\rho < 0$. 
I.e., $\rho=0$ corresponds to a one-way spacelike throat (of finite nonzero areal radius $|a|$)
that connects two SSS-AF BH regions: the region $\rho\in(-\infty,0)$ with $\rho\in(0,\infty)$ and hence avoiding the region $\sqrt{\rho^{2}+a^{2}}=0$ where the curvature singularity is formed. The above describes a novel type of regular black hole known as a black bounce.

The first black bounce model was proposed by
Simpson and Visser in their paper \cite{Simpson2019}. This 
model smoothly interpolates between the Schwarzschild black hole and the traversable Ellis Wormhole  
and is described by the static, spherically symmetric and asymptotically flat (SSS-AF) metric given by    
\begin{equation}\label{B_bounce}
\boldsymbol{ds^{2}} = - \left( 1 - \frac{2m}{\sqrt{\rho^{2} + a^{2}}} \right)\boldsymbol{dt^{2}} +  \left( 1 - \frac{2m}{ \sqrt{\rho^{2} + a^{2} } } \right)^{-1} \boldsymbol{d\rho^{2}}
+ \left( \rho^{2} + a^{2} \right)\boldsymbol{d\Omega^{2}},   
\end{equation}
being that $m$ and $a$ are real parameters;  
$t$ and $\rho$ are the temporal and radial\footnote{In the reference \cite{Simpson2019} the $\rho$ coordinate was called $r$.}
coordinates, respectively; while $\left( \rho^{2} + a^{2} \right)\boldsymbol{d\Omega^{2}}$, being $\boldsymbol{d\Omega^{2}}= \boldsymbol{d\theta^{2}} + \sin^{2}\theta \boldsymbol{d\varphi^{2}}$, is the line element of a two-dimensional sphere of radius $\sqrt{\rho^{2} + a^{2}}$.
The ranges of the coordinates 
are $t\in(-\infty,\infty)$, $\theta\in[0,\pi]$, and $\varphi\in[0,2\pi)$, whereas the $\rho$-coordinate as long as $a\neq0$ has range $\rho\in(-\infty,\infty)$.
Furthermore, in \cite{Simpson2019}, the spacetime metric (\ref{B_bounce}) was introduced 
as a minimalist modification of the ordinary Schwarzschild spacetime, such that when adjusting the parameters $a$ and $m$, this metric represents either: the ordinary Schwarzschild spacetime if $a=0\neq m$; whereas if $a\neq0$, depending on the relation between $m$ and $a$, it is either  
a traversable wormhole in the Morris-Thorne sense (if $2m<|a|$, which implies the absence of event horizons);
or becomes a one-way wormhole geometry with an extremal null throat (if $2m=|a|$, which implies the absence of traversable wormhole throat);
whereas it becomes a regular black hole of the black bounce type (if $2m>|a|$, which implies the existence of event horizons at $\rho=\pm\sqrt{4m^{2}-a^{2}}$ one on each side $\rho>0$, or $\rho<0$, of the bounce). Specifically, in this case, since for the region $\rho\in(-\sqrt{4m^{2}-a^{2}},\sqrt{4m^{2}-a^{2}})$ the $n(\rho)$-function is negative-definite, it follows that there the metric (\ref{B_bounce}) describes a Kantowski-Sachs type cosmology, with $\rho$ a temporal coordinate and $t$ a radial coordinate, being that 
$\rho = 0$ is a spacelike hypersurface that corresponds to a bounce in the time-dependent quantity $r(\rho)=\sqrt{\rho^{2} + a^{2}}$, which is in fact one of the scale factors of a Kantowski-Sachs cosmological model inside a black hole.
Here, it is important to emphasize that a black bounce is also a necessary feature of all black-universe
models \cite{Bronnikov2006,Bronnikov2007,Bronnikov2010,Bronnikov2012}, that is, nonsingular black holes that contain an expanding asymptotically isotropic cosmology beyond the horizon.

Nowadays, a growing interest in generalizing the Simpson-Visser black-bounce model has arisen: black-bounce Kerr model \cite{rotating_Bb}; black-bounce Reissner-Nordstr\"{o}m model \cite{Charged_rotating_Bb}; black-bounce Kerr-Newman model \cite{Charged_rotating_Bb}; new black-bounce spacetime model \cite{Simpson_4}. 

In this work, several black-bounce solutions in the context of Einstein-nonlinear electrodynamics gravity coupled to a self-interacting scalar field (GR-NLED-SF) are discussed. 
Furthermore, we obtain a novel black-bounce solution that does not belong to the Simpson-Visser type.

This paper is structured as follows. In the next section the canonical metric of a static spherically symmetric asymptotically flat black bounce is discussed. 
%
In Sec. \ref{BBs-ST-NLED} the field equations for the GR-NLED-SF are derived; we show that the line element for the Simpson-Visser black bounce is a pure magnetic exact solution of a GR-NLED-SF model; also, two novel black-bounce metrics of the Simpson-Visser type are presented as pure magnetic exact solutions of GR-NLED-SF; and some features of the solutions in the GR-NLED-SF frame are discussed.
In Sec. \ref{BB_non_Ellis} we obtain a novel black-bounce solution, which is not of the Simpson-Visser type, since does not admit the Ellis wormhole solution as a particular case. 
Final conclusions are given in the last Section. In this paper we use units where $G = k_{_{B}} = c =\hbar = 1$, and the metric signature $(-+++)$ is used throughout.


\section{The canonical metric of a static spherically symmetric asymptotically flat black-bounce spacetime: beyond the Simpson-Visser models } 
\label{method}

In what follows, we will describe a simple way to combine two SSS-AF spacetime geometries, 
one of the traversable wormhole (T-WH) type with one of the 
BH type, in order that a novel spacetime geometry of the SSS-AF black-bounce type be generated. \\

{\bf Method to generate black-bounce models:} Given a generic SSS-AF T-WH metric, 
\begin{equation}\label{TWH_MT}
\boldsymbol{ds^{2}}_{_{W\!H}} = - e^{2\Phi(r)}\boldsymbol{dt^{2}} + \frac{\boldsymbol{dr^{2}}}{ 1 - \frac{b(r)}{r}} + r^{2}\boldsymbol{d\Omega^{2}},
\end{equation}
where $\Phi(r)$ and $b(r)$ are smooth functions\footnote{I.e. $\Phi(r)$ and $b(r)$ are functions of class $\mathcal{C}^{\infty}$ for $r>0$.} on $r>0$, respectively known as redshift and shape functions, with wormhole throat radius, $r_{0}=|a|\!\in\mathbb{R}^{+}\!-\!\{0\}$, 
defined by $b(r_{0}) = r_{0}$, and for the which the T-WH properties
\begin{eqnarray} 
&&\textup{{\bf Wormhole domain:}}\hspace{2.8cm}   
 1 - \frac{b(r)}{r} \geq 0  \quad\quad\quad\quad \forall\quad\!\!\!\!\! r \geq |a| 
 \label{TWC1}\\
&& \textup{{\bf Absence of horizons:}}\hspace{2.5cm}  e^{2\Phi(r)}\in\mathbb{R}^{+}\!-\!\{0\}  \quad\quad  \forall\quad\!\!\!\!\! r\geq |a| 
\quad\quad \textup{ and } \quad\quad \Phi(r\rightarrow\infty)=0\label{TWC2}\\
&&\textup{{\bf Flaring out condition:}}\hspace{2.3cm} b'(r)\Big|_{r=|a|}<1,\label{TWC3}
\end{eqnarray}
with $'$ denoting derivative with respect to $r$,
are satisfied (see \cite{morris88,morris88-2,WEC} for details)
and given a generic SSS-AF-BH metric, 
\begin{equation}\label{BH_generic}
\boldsymbol{ds^{2}}_{_{B\!H}} = - \left( 1 - \frac{2 \mathcal{M}(r)}{r} \right)e^{2\delta(r)}\boldsymbol{dt^{2}} + \left( 1 - \frac{2 \mathcal{M}(r) }{r} \right)^{-1}\boldsymbol{dr^{2}} + r^{2}\boldsymbol{d\Omega^{2}},  \end{equation}
with $\delta(r)$ and $\mathcal{M}(r)$ smooth function\footnote{I.e. $\delta(r)$ and $\mathcal{M}(r)$ are functions of class $\mathcal{C}^{\infty}$ for $r>0$.} on $r>0$, with event horizon radius, $r=r_{h}\neq0$, such that
\begin{equation}\label{M_BB}
\mathcal{M}(r_{h}) = \frac{r_{h}}{2}, \quad\quad \mathcal{M}(r) < \frac{r}{2} \quad\quad \forall\quad\!\!\!\!\! r > r_{h}, \quad\quad  
\mathcal{M}(r \rightarrow \infty) = \textup{constant} \geq0  
\quad \textup{and}\quad 
\delta(r\rightarrow\infty)=0.
\end{equation}
In addition, as long as 
\begin{equation}\label{Non_Lorent}
1 - \frac{b(r)}{r}  < 0  \quad\quad\quad\quad\quad \forall\quad\!\!\!\!\! r \in[0, |a|)    
\end{equation}
a SSS-AF black-bounce model can be established by 
\begin{equation}\label{generalBB}
\boldsymbol{ds^{2}}_{_{B\!B}} \!=\! - \!\left(\!1 - \frac{2 \mathcal{M}(r)}{r}\!\right)\!e^{\Psi(r)}\boldsymbol{dt^{2}} + \frac{\boldsymbol{dr^{2}}}{ \left(1 - \frac{b(r)}{r}\right)\!\!\left( 1 - \frac{2 \mathcal{M}(r) }{r} \right)} + r^{2}\boldsymbol{d\Omega^{2}}, \quad\textup{with}\quad \Psi(r)=2\delta(r)+2\Phi(r) 
\end{equation}
which by construction admits Lorentz signature only at $r\geq b(r_{0})$, i.e. for $r\geq|a|$, whereas at $ r < b(r_{0})$, i.e. for $r<|a|$,  the metric becomes a non-Lorentzian metric, implying that the region $0 \leq r < |a|$ is not part of spacetime. 
In other words, this metric only admits physical interpretation in the region $r\in[|a|,\infty)$, while 
in the region $r\in(0,|a|)$ the metric   
suffers an unacceptable signature change, i.e. in the region $r\in(0,|a|)$ the metric signature
could be either $(++++)$ or $(--++)$, depending of the behavior of $\mathcal{M}(r)$. 
In the metric theories of gravity [as GR, $f(R)$ gravity, scalar-tensor theories, conformal gravity, for instance] the spacetime is modeled by a manifold with a metric of Lorentz signature  
at any point of the spacetime manifold. Thus, a region with a signature different from Lorentz signature would have no physical interpretation. Usually, the subset on which the signature changes to an unphysical one is, in one sense or another, an ``edge" of the manifold or of the allowed coordinate patch \cite{Plebanski}.
Therefore for the line element (\ref{generalBB}) the curvature singularity at $r=0$ is nonpathological since the region $r\in(0,|a|)$ lacks of physical interpretation. 
 
 In consequence, the radial coordinate $r$ has a range that increases from a minimum value at $r=|a|$ to $r\rightarrow\infty.$
Thus, in the spacetime geometry (\ref{generalBB}), the $r$-coordinate has a special geometric significance, where $4\pi r^{2}$ is the area of a sphere of radius $r$ centered at the origin, being that the origin of symmetries (staticity and spherical) is the two-dimensional spheres of radius $r=|a|$.
On the other hand, for the metric (\ref{generalBB}) the curvature invariants $R$, $R_{\alpha\beta}R^{\alpha\beta}$, and $R_{\alpha\beta\mu\nu}R^{\alpha\beta\mu\nu}$ are given by (\ref{Ext_R_BB_gen})$-$(\ref{Ext_RR4_BB_gen}),
indicating that all of them are regular everywhere except at $r=0$, and the reason is
because $b=b(r)$, $\Psi=\Psi(r)$, $\mathcal{M}=\mathcal{M}(r)$ are functions of class $\mathcal{C}^{\infty}$ for all $r>0$. 
However, since the region $0\leq r<|a|$ is not part of spacetime 
yields that the curvature invariants (\ref{Ext_R_BB_gen})$-$(\ref{Ext_RR4_BB_gen}) are well defined 
in the whole Lorentzian range\footnote{I.e., the curvature of spacetime is regular in all spacetime geometry.} $r\geq |a|$. Therefore, we conclude that the geometry (\ref{generalBB}), with  $b(r)$, $\mathcal{M}(r)$, and $\Psi(r)$  nontrivial functions that satisfy    
(\ref{TWC1})$-$(\ref{TWC3}), (\ref{M_BB}), and (\ref{Non_Lorent}), describes a spacetime with curvature regular everywhere. Finally, the metric (\ref{generalBB}) written in terms of the radial bounce coordinate $\rho$, defined as $r = \sqrt{\rho^{2} + a^{2}} \geq |a|$, takes the form: 
\begin{equation}\label{generalBB_BBC}
\boldsymbol{ds^{2}}_{_{B\!B}} = - \left( 1 - \frac{2 \mathcal{M}(r)}{\sqrt{ \rho^{2} + a^{2} } } \right)e^{\Psi(r)}\boldsymbol{dt^{2}} +  
\frac{ \rho^{2}\quad\!\!\!\!\boldsymbol{d\rho^{2}} }{ \left( \rho^{2} + a^{2} \right)\!\!\left( 1 - \frac{b(r)}{\sqrt{ \rho^{2} + a^{2} } } \right)\!\!\left( 1 - \frac{2 \mathcal{M}(r)}{\sqrt{ \rho^{2} + a^{2} } } \right) } + (\rho^{2}+a^{2})\boldsymbol{d\Omega^{2}}.  
\end{equation}

The determinant of this metric is given by  
\begin{equation}\label{detG}
g=det(g_{\alpha\beta})  
= - \frac{ \rho^{2}  \left(\rho^{2} + a^{2}\right)^{\frac{3}{2}} e^{\Psi(r)} }{ \left[ \sqrt{ \rho^{2} + a^{2} } - b(r) \right] }.  
\end{equation}
Since (\ref{TWC1})  
and the requirement that $\Psi$ is a function of class $\mathcal{C}^{\infty}$ for all $r > |a|$, or what is the same,  
$\Psi\in\mathcal{C}^{\infty}$ for all real values of $\rho$ defined by $r=\sqrt{\rho^{2} + a^{2}}\geq|a|$ yields that the determinant (\ref{detG})  
is finite nonzero and negative definite for all $\rho\in(-\infty,\infty)$. In particular for $\rho=0$ this determinant\footnote{Note that $\frac{\frac{d\rho^{2}}{d\rho}}{\frac{d}{d\rho}\left[ \sqrt{ \rho^{2} + a^{2} } - b(r)  \right] } = \frac{2 \sqrt{ \rho^{2} + a^{2} } }{ 1 -  b'(r) },$ implying  $\lim\limits_{\rho \to 0}\frac{\rho^{2} }{ \sqrt{ \rho^{2} + a^{2} } - b(r) }  = \frac{2|a|}{ 1 -  b'(|a|) }.$ } reduces to $\lim\limits_{\rho \to 0} det(g_{\alpha\beta})  = - \frac{2a^{4}e^{\Psi(|a|)}}{ 1 -  b'(|a|) }  \in\mathbb{R}^{-}\!-\!\{0\}$ as long as (\ref{TWC3}). 
Hence, the metric (\ref{generalBB_BBC}), for the case with $a\neq0$, is well defined for all $\rho\in(-\infty,\infty)$. \\
Therefore, the spacetime described by metric (\ref{generalBB_BBC}) $\equiv$ (\ref{generalBB}) has been carefully designed
to be a minimalist modification of the ordinary
black hole spacetime (\ref{BH_generic}); when adjusting 
$a$, $r_{h}$, $\mathcal{M}=\mathcal{M}(r)$ and $b=b(r)$,
this metric represents either:
%
%
\begin{itemize}
%
    \item[i)] If ($\mathcal{M}=0\neq b$) becomes a traversable wormhole of the Morris-Thorne type, with throat of radius $r_{0}=|a|$ located at $\rho_{0}=0$ (in the radial bounce coordinate $\rho$).  
     \item[ii)] If ($\mathcal{M}\neq 0 \neq b$), 
     such that $b(r_{0})>2\mathcal{M}(r_{h})$ (which implies absence of horizon), becomes 
     a traversable wormhole of the Morris-Thorne type, with throat of radius $r_{0}=|a|$ located at $\rho_{0}=0$.
    \item[iii)] If ($\mathcal{M}\neq 0 \neq b$), 
     such that $b(r_{0}) = 2\mathcal{M}(r_{h})$ i.e. $r_{0}=r_{h}$ (which implies absence of a T-WH throat) becomes a one-way wormhole with a null throat of radius $r_{0}=|a|$ located at $\rho_{0}=0$.  
    \item[iv)] If ($\mathcal{M}\neq 0 = b$) becomes an ordinary black hole spacetime with event horizon radius given by  $r_{h}=2\mathcal{M}(r_{h})$.    
    \item[v)] If ($\mathcal{M}\neq 0 \neq b$), such that $b(r_{0}) < 2\mathcal{M}(r_{h})$, becomes a regular black hole (black bounce) with a one-way spacelike throat at $\rho_{0}=0$ (of radius $r_{0}=|a|$),  
    and with two (outer) event horizons of radio 
    $r_{h}=2\mathcal{M}(r_{h})>|a|$, located at 
    $\rho_{h} = \pm\sqrt{r^{2}_{h} - a^{2}}\neq0$.  
    %
\end{itemize}

{\bf Particular case: black bounces of the Simpson-Visser type.} 
For the case $b(r) = a^{2}/r$ with $\Psi(r)=2\delta(r)+2\Phi(r)=0$  
the line element (\ref{generalBB}) takes the form: 
\begin{equation}\label{BB_structure}
\boldsymbol{ds^{2}}_{_{B\!B}} = - \left( 1 - \frac{2 \mathcal{M}(r)}{r} \right)\boldsymbol{dt^{2}} + \frac{\boldsymbol{dr^{2}}}{\Big( 1 - \frac{a^{2}}{r^{2}} \Big)\Big( 1 - \frac{2 \mathcal{M}(r) }{r} \Big) } + r^{2}\boldsymbol{d\Omega^{2}},     
\end{equation}
whereas in terms of the radial $\rho$-coordinate ($\rho=\pm\sqrt{r^{2}-a^{2}}$), it becomes 
\begin{equation}\label{generalBB_BBC_SV}
\boldsymbol{ds^{2}}_{_{B\!B}} = - \left( 1 - \frac{2 \mathcal{M}(r)}{\sqrt{ \rho^{2} + a^{2} } } \right)\boldsymbol{dt^{2}} +  
\frac{ \boldsymbol{d\rho^{2}} }{ \left( 1 - \frac{2 \mathcal{M}(r)}{\sqrt{ \rho^{2} + a^{2} } } \right) } + (\rho^{2}+a^{2})\boldsymbol{d\Omega^{2}}. 
\end{equation}
For this case the determinant (\ref{detG}) becomes
\begin{equation}\label{detSV}
g=det( g_{\alpha\beta} ) = - \left(\rho^{2} + a^{2}\right)^{2},    
\end{equation}
which is finite nonzero and negative definite 
for all $\rho$ real.
The spacetime metric (\ref{generalBB_BBC_SV}) $\equiv$ (\ref{BB_structure}) 
describes a generic black-bounce geometry of the Simpson-Visser type. In particularly, this type of black-bounce spacetimes are such that for $a\neq0=\mathcal{M}(r)$ becomes the traversable Ellis WH metric presented originally in \cite{Ellis}.  Adjusting  
$a$, $r_{h}$ and  $\mathcal{M}=\mathcal{M}(r)$, this metric represents either: 
\begin{itemize}
%
    \item[i)] The ordinary traversable Ellis wormhole spacetime if ($\mathcal{M}=0\neq a$). 
    \item[ii)] A traversable wormhole in the Morris-Thorne sense if ($\mathcal{M}\neq0\neq a$ with $|a| > r_{h}$).
    \item[iii)] An one-way wormhole with a null throat if ($\mathcal{M}\neq0\neq a$ with $|a| = r_{h}$).
    \item[iv)] An ordinary black hole spacetime if ($\mathcal{M}\neq 0 = a$).   
    \item[v)] A black bounce with a one-way spacelike throat if ($\mathcal{M}\neq 0 \neq a$ with $r_{h} > |a|$).
\end{itemize}

\section{Field equations for a generic static spherically symmetric, purely magnetic spacetime configuration}
\label{BBs-ST-NLED}

The GR-NLED-SF theory is defined by the following action:
\begin{equation}\label{action_STNLED}
S[g_{ab},\phi,A_{c}] = \int d^{4}x \sqrt{-g} \left\{ \frac{1}{16\pi}\left(R - \frac{1}{2}\partial_{\mu}\phi\partial^{\mu}\phi  - 2 \mathscr{U}\!(\phi) \right) + \frac{1}{4\pi}\mathcal{L}(\mathcal{F}) \right\},
\end{equation}
where $R$ is the scalar curvature, $\phi$ is a scalar field which is minimally coupled to gravity, $\mathscr{U}=\mathscr{U}(\phi)$ is the scalar potential, whereas $\mathcal{L}=\mathcal{L}(\mathcal{F})$ is a function of the electromagnetic invariant $\mathcal{F}\equiv \frac{1}{4}F_{\alpha\beta}F^{\alpha\beta}$, being $F_{ab}=2\partial_{[a}A_{b]}$  
the components of the electromagnetic field tensor $\boldsymbol{F}=\frac{1}{2}F_{\alpha\beta} \boldsymbol{dx^{\alpha}} \wedge \boldsymbol{dx^{\beta}}$ and $A_{a}$ are the components of the electromagnetic potential. 

The GR-NLED-SF field equations arising from action (\ref{action_STNLED}) are 
\begin{equation}\label{ES_NLED_Eqs}
G_{\alpha}{}^{\beta} = 8\pi (E_{\alpha}{}^{\beta})\!_{_{_{S \! F}}} \!+ \! 8\pi (E_{\alpha}{}^{\beta})\!_{_{_{N \! L \! E \! D}}}, \quad\quad\quad 
\nabla_{\mu}(\mathcal{L}_{_{\mathcal{F}}}F^{\mu\nu}) = 0 = d\boldsymbol{F},  
\quad\quad\quad \nabla^{2}\phi = 2\hspace{0.03cm}\dot{\mathscr{U}}, %
\end{equation}
where, $\mathcal{L}_{_{\mathcal{F}}}\equiv \frac{d\mathcal{L}}{d\mathcal{F}}$ and $\dot{\mathscr{U}} = \frac{d\mathscr{U}}{d\phi}$, whereas $G_{\alpha}{}^{\beta} = R_{\alpha}{}^{\beta} - \frac{R}{2} \delta_{\alpha}{}^{\beta}$ denotes the components of the Einstein tensor, $(E_{\alpha}{}^{\beta})\!_{_{_{S \! F}}}$ are the components of the energy-momentum tensor of self-interacting scalar field, 
\begin{equation}\label{E_SF}
8 \pi(E_{\alpha}{}^{\beta})\!_{_{_{S \! F}}} = -\frac{1}{4}(\partial_{\mu}\phi \hspace{0.04cm} \partial^{\mu}\phi)\delta_{\alpha}{}^{\beta} + \frac{1}{2}\partial_{\alpha}\phi \hspace{0.04cm} \partial^{\beta}\phi - \mathscr{U} \hspace{0.03cm} \delta_{\alpha}{}^{\beta},     
\end{equation}
whereas, $(E_{\alpha}{}^{\beta})\!_{_{_{N \! L \! E \! D}}}$  are the components of the NLED energy-momentum tensor,
\begin{equation}\label{NLED_EM}
8\pi(E_{\alpha}{}^{\beta})\!_{_{_{N \! L \! E \! D}}} = 2\mathcal{L}_{_{\mathcal{F}}}F_{\alpha\mu}F^{\beta\mu} - 2\mathcal{L}\hskip.06cm\delta_{\alpha}{}^{\beta}. 
\end{equation}
Our aim is to find a solution of the set of Eqs. (\ref{ES_NLED_Eqs}),
that describes a SSS-AF
charged black-bounce solution with a nontrivial scalar field. Therefore, we will assume that the scalar field is static and spherically symmetric, $\phi = \phi(r)$, and also that the metric takes the static and spherically symmetric form
\begin{equation}\label{SSSmet}
ds^{2} =  - e^{ A(r) }dt^{2} + e^{ B(r) }dr^{2}  + r^{2}(d\theta^{2}  + \sin^{2}\theta d\varphi^{2}),
\end{equation}
with $A=A(r)$ and $B=B(r)$ being unknown functions depending only on $r$.

Below, we include the explicit form of the field equations assuming both the SSS  
for the metric (\ref{SSSmet}), SSS scalar field $\phi(r)$, and an arbitrary NLED $\mathcal{L}(\mathcal{F})$ model.
%
%
For a generic SSS spacetime metric ansatz (\ref{SSSmet})  
the non-null components of the Einstein tensor are given by
%
\begin{equation}\label{GabSSS}
G_{t}{}^{t}\!=\!\frac{ e^{^{\!\!-B}} }{ r^{2} }\!\!\left( -rB' - e^{^{\!B}} + 1 \right)\!, \hspace{0.25cm}  G_{r}{}^{r} \!=\! \frac{ e^{^{\!\! -B}} }{ r^{2} }\!\!\left( rA' - e^{^{\!B}} + 1 \right)\!, \hspace{0.25cm} G_{\theta}{}^{\theta}\!=\! G_{\varphi}{}^{\varphi}\!=\!\frac{ e^{^{\!\!-B}} }{ 4r }\!\!\left( rA'^{2} - rA'B' + 2rA'' + 2A' - 2B' \right)\!.
\end{equation}
%
The nontrivial components of the energy-momentum tensor of self-interacting scalar field are
\begin{equation}\label{EttyErr}
8\pi E_{t}{}^{t} = 8\pi E_{\theta}{}^{\theta} = 8\pi E_{\varphi}{}^{\varphi}   = -\frac{1}{ 4 } e^{ -B} \phi'^{2}  - \mathscr{U}, \quad\quad\quad\quad 8\pi E_{r}{}^{r} = \frac{1}{ 4 } e^{ -B} \phi'^{2} - \mathscr{U}. \end{equation}
Regarding the electromagnetic field tensor, since the spacetime is SSS, we can restrict ourselves to purely magnetic field; i.e., $\mathcal{E} = 0$ and $\mathcal{B} \neq 0$,
thus the electromagnetic field tensor has the form $F_{\alpha\beta} = \mathcal{B}\left( \delta^{\theta}_{\alpha}\delta^{\varphi}_{\beta} - \delta^{\varphi}_{\alpha}\delta^{\theta}_{\beta} \right).$ 
In this way, for a  
SSS spacetime with line element (\ref{SSSmet}), the general solution of the equations $\nabla_{\mu}(\mathcal{L}_{_{\mathcal{F}}}F^{\mu\nu})=0$ is given by
\begin{equation}\label{fabSOL}
F_{\theta\varphi} =  r^{4} \mathcal{Q}(r) \sin\theta.
\end{equation}
Then, $\boldsymbol{F} = r^{4} \mathcal{Q}(r)\sin\theta \hspace{0.1cm} \boldsymbol{d\theta} \wedge \boldsymbol{d\varphi}$, therefore $d\boldsymbol{F} = 0 = (r^{4} \mathcal{Q}(r))' \sin\theta \hspace{0.1cm} \boldsymbol{dr} \wedge \boldsymbol{d\theta} \wedge \boldsymbol{d\varphi}$, yields $\mathcal{Q}(r) = \sqrt{2}\hskip.06cm q/r^{4}$, where
$\sqrt{2}\hskip.06cm q$ is an integration constant, in which it plays the role of the magnetic charge.
Hence, the components of the electromagnetic field tensor, and the invariant $\mathcal{F}$ are
respectively given by
\begin{equation}\label{magnetica}
F_{\alpha\beta} = \sqrt{2} \hspace{0.05cm} q  \sin \theta \hspace{0.05cm} \left( \delta^{\theta}_{\alpha}\delta^{\varphi}_{\beta} - \delta^{\varphi}_{\alpha}\delta^{\theta}_{\beta} \right),  \quad\quad\quad\quad \mathcal{F} =  \frac{q^{2}}{r^{4}}   
\end{equation}

Finally, the  energy-momentum tensor components 
for NLED, assuming the SSS spacetime with metric (\ref{SSSmet}), the purely magnetic field  (\ref{magnetica}), and a generic Lagrangian density $\mathcal{L}(\mathcal{F})$, are given by 
\begin{eqnarray}\label{E_nled}
8\pi (E_{t}{}^{t})\!_{_{_{N \! L \! E \! D}}} = 8\pi (E_{r}{}^{r})\!_{_{_{N \! L \! E \! D}}} =  -2\mathcal{L}, 
\quad\quad\quad 8\pi (E_{\theta}{}^{\theta})\!_{_{_{N \! L \! E \! D}}} = 8\pi (E_{\varphi}{}^{\varphi})\!_{_{_{N \! L \! E \! D}}} =  2(2\mathcal{F}\mathcal{L}_{\mathcal{F}} - \mathcal{L}). 
\end{eqnarray}
%
Inserting the above given components in the field equations written as
$C_{\alpha}{}^{\beta} = G_{\alpha}{}^{\beta} - 8\pi [ (E_{\alpha}{}^{\beta})\!_{_{_{S \! F }}} + (E_{\alpha}{}^{\beta})\!_{_{_{N \! L \! E \! D}}} ] = 0$, we obtain that the GR-NLED-SF field equations for the metric ansatz (\ref{SSSmet}) and the magnetic field (\ref{magnetica}) take the form:
\begin{eqnarray}
&&\!C_{t}{}^{t}\!=\!0\hspace{1.4cm}\!\Rightarrow\!\hspace{0.4cm}
\frac{ e^{^{\!\!-B}} }{ r^{2} }\!\!\left( -rB' - e^{^{\!B}} + 1 \right) + \frac{1}{ 4 } e^{ -B} \phi'^{2}  + \mathscr{U} + 2\mathcal{L} =0,
\label{Eqt}\\
&&\nonumber\\
&& \!C_{r}{}^{r}\!=\!0\hspace{1.4cm}\!\Rightarrow\!\hspace{0.4cm} \frac{ e^{^{\!\! -B}} }{ r^{2} }\!\!\left( rA' - e^{^{\!B}} + 1 \right) - \frac{1}{ 4 } e^{ -B} \phi'^{2} + \mathscr{U} + 2\mathcal{L} = 0,
\label{Eqr}\\
&&\nonumber\\
&& \!C_{\theta}{}^{\theta}\!=\!C_{\varphi}{}^{\varphi}\!=\!0\hspace{0.4cm}\!\Rightarrow\!\hspace{0.4cm} 
\frac{ e^{^{\!\!-B}} }{ 4r }\!\!\left( rA'^{2} - rA'B' + 2rA'' + 2A' - 2B' \right) + \frac{1}{ 4 } e^{ -B} \phi'^{2}  + \mathscr{U} - 2(2\mathcal{F}\mathcal{L}_{\mathcal{F}} - \mathcal{L})  =0,
\label{Eqte}
\end{eqnarray}
whereas the scalar field should satisfy
\begin{equation}\label{phi2}
2r\phi'' + (4 + rA' - rB')\phi' - 4r e^{B}\dot{\mathscr{U}} = 0.  
\end{equation}

\subsection{Solution for nontrivial scalar field, vanishing scalar potential, and vanishing  electromagnetic field:\\Ellis wormhole metric}

For the case $a\neq0$, $\mathscr{U}=F_{\alpha\beta}=\mathcal{L}(\mathcal{F})=0$, the field equations (\ref{Eqt})$-$(\ref{phi2})   
are solved by the following static and spherically symmetric line element:
\begin{equation}\label{EllisWH}
ds^{2}_{_{W\!H}} =  - dt^{2} + \frac{dr^{2}}{ 1 - \frac{a^{2}}{r^{2}} }  + r^{2}(d\theta^{2}  + \sin^{2}\theta d\varphi^{2}),
\end{equation}
with a imaginary scalar field, given by 
\begin{equation}\label{Ellis_SF}
\phi(r) = 2i\tan^{^{\!\!-1}}\!\!\left( \sqrt{\frac{r^{2} - a^{2}}{a^{2}}}\right).
\end{equation}
The metric (\ref{EllisWH}), originally introduced in \cite{Ellis}, admits a T-WH interpretation since satisfies the properties (\ref{TWC1})-(\ref{TWC3}), and is known as the Ellis wormhole metric.\\
Indeed, defining a new scalar field by $\psi = i\phi$ (phantom field), and using $\mathscr{U}=\mathcal{L}(\mathcal{F})=0$, the action (\ref{action_STNLED}) takes the form  
\begin{equation}\label{actionL2}
S[g_{ab},\psi] = \int d^{4}x \sqrt{-g} \left\{ \frac{1}{16\pi}\left(R + \frac{1}{2}\partial_{\mu}\psi\partial^{\mu}\psi \right)  \right\}.
\end{equation}
This gravitational action defines a theory that admits the Ellis WH metric 
as an exact solution and  with $\psi$ given by
\begin{equation}
\psi =  2\tan^{-1}\left( \sqrt{ \frac{r^{2}  - a^{2}}{a^{2}} } \right)  \in \mathbb{R}.
\end{equation}
This is the action that was used by Ellis in Ref. \cite{Ellis} to get the wormhole solution (\ref{EllisWH}).
%

\subsection{Simpson-Visser black bounce as a pure magnetic exact solution of the GR-NLED-SF field equations}    
The following NLED-SF theory defined by a scalar potential and a power-law Maxwell NLED model, given respectively by
\begin{equation}\label{U_L_SV}
\mathscr{U}(\phi) = \mathscr{U}_{_{0}}\hspace{0.04cm}\cosh^{5}\left(\frac{\phi}{2}\right), %
\quad\quad\quad 
\mathcal{L}(\mathcal{F}) = s_{_{0}}|\mathcal{F}|^{^{\!\frac{5}{4}}}.
\end{equation}
where $\mathscr{U}_{_{0}}$ and $s_{_{0}}$ are real parameters of the theory, such that
for the case $\mathscr{U}_{_{0}}= 4\hspace{0.04cm}m/(5\hspace{0.04cm}|q|^{3})$ and $s_{_{0}}=3m/(5\sqrt{|q|})$ being that $q$ and $m$ are real parameters, defines a NLED-SF model for the which the metric  
\begin{equation}\label{BH_WH_Schw_Ellis}
\boldsymbol{ds^{2}}_{_{B\!B}} = - \left( 1 - \frac{2m}{r} \right)\boldsymbol{dt^{2}} + \frac{\boldsymbol{dr^{2}}}{\Big( 1 - \frac{q^{2}}{r^{2}} \Big)\Big( 1 - \frac{2m}{r} \Big) }
+ r^{2}\boldsymbol{d\Omega^{2}}   
\end{equation}
together with the scalar field 
\begin{equation}\label{sf_SV}
\phi(r) = 2i\tan^{^{\!\!-1}}\!\!\left( \sqrt{\frac{r^{2} - q^{2}}{q^{2}}}\right),    
\end{equation}
is a pure magnetic exact solution of the GR-NLED-SF field equations (\ref{Eqt})$-$(\ref{phi2}), being that $q$ is the magnetic charge. 

The metric (\ref{BH_WH_Schw_Ellis}) describes a black bounce model of the type (\ref{BB_structure}), which according with our method to produce black-bounce models,  
come from considering the Schwarzschild metric
\begin{equation}
\boldsymbol{ds^{2}}_{_{B\!H}} = - \left( 1 - \frac{2m}{r} \right)\boldsymbol{dt^{2}} + \left( 1 - \frac{2m}{r} \right)^{-1}\boldsymbol{dr^{2}} + r^{2}\boldsymbol{d\Omega^{2}},   
\end{equation}
as the SSS-AF-BH metric (\ref{BH_generic}), and the Ellis solution (\ref{EllisWH}) with $a^{2} = q^{2}$ as the T-WH metric (\ref{TWH_MT}).

Finally, if we change the radial coordinate to $\rho^{2} = r^{2} - q^{2}$, which implies $\boldsymbol{d\rho} = \pm (r/\sqrt{r^{2} - q^{2}})\boldsymbol{dr}$, the line element (\ref{BH_WH_Schw_Ellis}) takes the form
\begin{equation}\label{BH_WH_2}
\boldsymbol{ds^{2}}_{_{B\!B}} = - \left( 1 - \frac{2m}{\sqrt{\rho^{2} + q^{2}}} \right)\boldsymbol{dt^{2}} +  \left( 1 - \frac{2m}{ \sqrt{\rho^{2} + q^{2} } } \right)^{-1}\boldsymbol{d\rho^{2}}
+ \left( \rho^{2} + q^{2} \right)\boldsymbol{d\Omega^{2}},   
\end{equation}
which corresponds to the original Simpson-Visser black-bounce model (\ref{B_bounce}) with $a^{2}=q^{2}$.  

{\bf Avoiding the Penrose singularity theorem:}  
Using (\ref{sf_SV}), (\ref{WEC_SF}), and (\ref{WEC_effect}) for the Simpson-Visser black-bounce spacetime, one obtains  
\begin{equation}
8\pi(\rho_{t})\!_{_{_{S \! F }}} + 8\pi (P_{r})\!_{_{_{S \! F }}} = 8\pi(\rho_{t})\!_{_{_{e\!f\!f}}} + 8\pi (P_{r})\!_{_{_{e\!f\!f}}} = -\frac{2q^{2}}{r^{4}}\!\left(1-\frac{2\mathcal{M}}{r}\right) < 0 \quad\quad\quad  \forall\quad\!\!\!\!\! 
r \in(2\mathcal{M},\infty), 
\end{equation}
which indicates that the local energy density of the scalar field $\rho_{\!_{_{SF}}}= (E_{\mu\nu})\!_{_{_{S\!F}}}k^{\mu}k^{\nu}$, and the local energy density of the total energy/mater source (self-interacting scalar field plus the nonlinear electromagnetic field) $\rho_{\!_{_{eff}}}= (E_{\mu\nu})\!_{_{_{e\!f\!f}}}k^{\mu}k^{\nu}$, for every timelike vector field $\boldsymbol{k}= k^{\mu}\partial_{\mu}$, are not positive definite in the whole spacetime,  
whereas the power-law Maxwell NLED (\ref{U_L_SV}) for the purely magnetic field (\ref{magnetica}) holds
\begin{equation}
\mathcal{F} = \frac{q^{2}}{r^{4}} >0, \quad\quad\quad \mathcal{L} = s_{_{0}}\mathcal{F}^{^{\frac{5}{4}}} > 0,  \quad\quad\quad \mathcal{L}_{\mathcal{F}} = \frac{5s_{_{0}}}{4}\mathcal{F}^{^{\frac{1}{4}}} > 0  \quad\quad \forall\quad\!\!\!\!\! r    
\end{equation}
according to (\ref{WEC_NLED}); this implies that the local energy density of the nonlinear electrodynamic field $\rho_{\!_{_{NLED}}}= (E_{\mu\nu})\!_{_{_{N\!L\!E\!D}}}k^{\mu}k^{\nu}$ 
is positive defined everywhere.
Thus, the Simpson-Visser black bounce as a pure magnetic exact solution of GR-NLED-SF  
does not satisfy the WEC, being that the self-interacting scalar field is only responsible for the violation of WEC, and therefore the gravitational collapse, from which the Simpson-Visser black bounce is produced, avoiding the singularity theorem of Penrose \cite{Penrose}. See Appendix \ref{Avoiding_singularity} for details.\\ 

\subsection{Canonical acoustic black bounce as a pure magnetic exact solution of the GR-NLED-SF field equations} 
The following NLED-SF theory defined by a scalar potential and a NLED model, given respectively by
\begin{equation}\label{NLED_SF_CABB}
\mathscr{U}(\phi) = \mathscr{U}_{_{1}}\hspace{0.04cm}\cosh^{8}\left(\frac{\phi}{2}\right), %
\quad\quad\quad 
\mathcal{L}(\mathcal{F})  
= s_{_{1}}\mathcal{F}^{^{2}} - \sqrt{ \frac{ 3 s_{_{1}} | \mathcal{F}|^{^{3}} }{ 2 a^{2} }  },
\end{equation}
where $a$, $\mathscr{U}_{_{1}}$, and $s_{_{1}}$  are real parameters of the theory, such that for the case 
$\mathscr{U}_{_{1}}= q^{2}/a^{6}$ and $s_{_{1}} = 3a^{2}/(2q^{2})$,  
defines a NLED-SF model for the which the metric,
\begin{equation}\label{CABH_BB}
\boldsymbol{ds^{2}}_{_{B\!B}} = - \left( 1 - \frac{q^{2}}{r^{4}} \right)\boldsymbol{dt^{2}} + \frac{\boldsymbol{dr^{2}}}{\left(1 - \frac{a^{2}}{r^{2}} \right)\left(1 - \frac{q^{2}}{r^{4}} \right)} + r^{2}\boldsymbol{d\Omega^{2}} 
\end{equation}
together with the scalar field,
\begin{equation}\label{SF_cabh}
\phi(r) = 2i\tan^{^{\!\!-1}}\!\!\left( \sqrt{\frac{r^{2} - a^{2}}{a^{2}}}\right),    
\end{equation}
is a pure magnetic exact solution of the GR-NLED-SF field equations (\ref{Eqt})$-$(\ref{phi2}), being that the parameter $q$ is the magnetic charge.\\
%
%
The metric (\ref{CABH_BB}) defines  
a black-bounce model of the Simpson-Visser type (\ref{BB_structure}), which smoothly interpolates between the Ellis wormhole metric (if $q=0\neq a$) and the  
canonical acoustic black hole (CABH) metric (if $q\neq0=a$) given by 
\begin{equation}\label{CABH}
\boldsymbol{ds^{2}}_{_{B\!H}} = - \left( 1 - \frac{q^{2}}{r^{4}} \right)\boldsymbol{dt^{2}} + \left(1 - \frac{q^{2}}{r^{4}} \right)^{-1}\boldsymbol{dr^{2}} + r^{2}\boldsymbol{d\Omega^{2}}, 
\end{equation}
which was originally derived in \cite{Visser}. The acoustic black holes (or sonic black holes) are acoustic analogues of the gravitational black holes.  
Specifically, an acoustic black hole forms when the velocity of the fluid exceeds the velocity of sound on some closed surface. That surface forms a sonic horizon, an exact sonic analog of a black hole horizon where the sound modes, or phonons (rather than light waves), cannot escape the event horizon (see \cite{Unruh} for details).
In the gravitational context, recently in \cite{BHs_CABH}, the line element (\ref{CABH}) was reinterpreted as an exact gravitational black hole solution of the Einstein-scalar-Gauss-Bonnet field equations.  

To put in context,   
as an application of the black-bounce models generator method, the black-bounce metric (\ref{CABH_BB}) is  
generated using the Ellis wormhole as the T-WH metric (\ref{TWH_MT}), and the CABH  
as the BH metric (\ref{BH_generic}). 

On the other hand, using the radial coordinate $\rho$, defined by $\rho^{2} = r^{2} - a^{2}$, the metric (\ref{CABH_BB}) takes the form 
\begin{equation}\label{BH_WH_CA}
\boldsymbol{ds^{2}}_{_{B\!B}} = - \left( 1 - \frac{q^{2}}{ \left( \rho^{2} + a^{2} \right)^{2} } \right)\boldsymbol{dt^{2}} +  \left( 1 - \frac{q^{2}}{\left( \rho^{2} + a^{2} \right)^{2} } \right)^{-1}\boldsymbol{d\rho^{2}}
+ \left( \rho^{2} + a^{2} \right)\boldsymbol{d\Omega^{2}},   
\end{equation}
being that the canonical acoustic black-bounce metric written in the bounce coordinates $(x^{\alpha})=(t,\rho,\theta,\phi)$ are defined as $t\in(-\infty,\infty)$, $\rho\in(-\infty,\infty)$, $\theta\in[0,\pi]$, and $\phi\in[0,2\pi)$. 

Hence, adjusting the parameters $q$ and $a$, the spacetime metric (\ref{BH_WH_CA}) $\equiv$ (\ref{CABH_BB})
admits the following interpretations: 
\begin{itemize}
%
    \item[i)] The ordinary traversable Ellis wormhole spacetime if ($q=0\neq a$). 
    \item[ii)] A new traversable Morris–Thorne wormhole (which generalizes to the Ellis solution WH) if ($q\neq0\neq a$ such that $|q| < a^{2}$) with WH throat at $\rho_{0} = 0$ of radius $r_{0}=|a|$. 
    \item[iii)] An one-way wormhole with a null throat if ($|q|=a^{2}\neq0$).
    \item[iv)] The ordinary canonical acoustic black hole spacetime if ($q\neq 0 = a$).   
    \item[v)] A black bounce if ($q\neq 0 \neq a$ such that $|q| > a^{2}$), with a one-way spacelike throat at $\rho_{0}=0$ (of radius $r_{0}=|a|$), and with event horizons at $\rho_{h}=\pm\sqrt{|q|-a^{2}}\neq0$  (of radius $r_{h}=\sqrt{|q|}>|a|$). 
\end{itemize}

On the other hand, using (\ref{SF_cabh}), (\ref{WEC_SF}), and (\ref{WEC_effect}), for the canonical acoustic black-bounce solution, yields 
\begin{equation}\label{CABH_nWEC_1}
8\pi(\rho_{t})\!_{_{_{S \! F }}} + 8\pi (P_{r})\!_{_{_{S \! F }}} = 8\pi(\rho_{t})\!_{_{_{e\!f\!f}}} + 8\pi (P_{r})\!_{_{_{e\!f\!f}}} = -\frac{2a^{2}}{r^{4}}\!\left(1-\frac{q^{2} }{r^{4}}\right) < 0 \quad\quad\quad  \forall\quad\!\!\!\!\! r \in(\sqrt{|q|},\infty),  
\end{equation}
whereas, the NLED model (\ref{NLED_SF_CABB}) for the purely magnetic field (\ref{magnetica}), $\mathcal{F} = \frac{q^{2}}{r^{4}} >0$,  holds,
\begin{equation}\label{CABH_nWEC_2}
\mathcal{L} = s_{_{1}}\mathcal{F}^{^{2}} - \sqrt{ \frac{ 3 s_{_{1}}  \mathcal{F}^{^{3}} }{ 2 a^{2} }  } = - \frac{3 q^{2} }{ 2 r^{6}}\left( 1 - \frac{a^{2}}{r^{2}} \right) < 0,  \quad \mathcal{L}_{\mathcal{F}} = 2s_{_{1}}\mathcal{F} - \sqrt{ \frac{ 27 s_{_{1}}  \mathcal{F} }{ 8 a^{2} }  } = -\frac{3}{r^{2}}\left( \frac{3}{4} - \frac{a^{2}}{r^{2}} \right) < 0  \quad \forall\quad\!\!\!\!\! r \in(|a|,\infty).     
\end{equation}
Then, according to  (\ref{WEC_SF}), (\ref{WEC_NLED}), and (\ref{WEC_effect}), the inequalities (\ref{CABH_nWEC_1}) and (\ref{CABH_nWEC_2}) imply the local energy densities $\rho_{\!_{_{SF}}}= (E_{\mu\nu})\!_{_{_{S\!F}}}k^{\mu}k^{\nu}$, $\rho_{\!_{_{NLED}}}= (E_{\mu\nu})\!_{_{_{N\!L\!E\!D}}}k^{\mu}k^{\nu}$,  and $\rho_{\!_{_{eff}}}= (E_{\mu\nu})\!_{_{_{e\!f\!f}}}k^{\mu}k^{\nu}$, associated respectively to scalar field, nonlinear electromagnetic field, and the total energy/matter source, 
are not positive defined in the whole canonical acoustic black-bounce spacetime. Hence, the canonical acoustic black bounce as a pure magnetic exact solution of GR-NLED-SF does not satisfy the WEC, and therefore the gravitational collapse that resulted in the canonical acoustic black bounce avoid the Penrose singularity theorem \cite{Penrose}.

\subsection{Black-bounce-4D-Einstein-Gauss-Bonnet as an exact solution of the ES-NLED field equations}

The following NLED-SF theory defined by a scalar potential and a NLED model, given respectively by
\begin{eqnarray}
\mathscr{U}(\phi(r)) &=&  \bigintss^{r}_{\!\!\!\!\!\!\!\!\!\!|a|}\hspace{0.3cm}\frac{ 2 a^{2} \left(  r^{3} + 2 \hspace{0.03cm}\alpha\hspace{0.03cm}q^{2} - \sqrt{ (r^{3} + 8\hspace{0.03cm}\alpha\hspace{0.03cm} q^{2})r^{3}} \right) }{ \alpha \sqrt{ (r^{3} + 8\hspace{0.03cm}\alpha\hspace{0.03cm}q^{2})r^{9}} } \hspace{0.1cm} dr,\\ 
\mathcal{L}(\mathcal{F}(r)) &=& - \frac{3}{4\hspace{0.03cm}\alpha} + \frac{a^{2}}{2\hspace{0.03cm}\alpha\hspace{0.03cm}r^{2}} + \frac{ 3r^{5} - 2a^{2}r^{3} + 12 \alpha q^{2} r^{2} - 4a^{2}\alpha q^{2}}{4\alpha \sqrt{ (r^{3} + 8\hspace{0.03cm}\alpha\hspace{0.03cm}q^{2})r^{7}} } -\!\!\! \bigintss^{r}_{\!\!\!\!\!\!\!\!\!\!|a|}\hspace{0.2cm}\frac{ a^{2}\!\left(  r^{3} + 2 \hspace{0.03cm}\alpha\hspace{0.03cm}q^{2} - \sqrt{ (r^{3} + 8\hspace{0.03cm}\alpha\hspace{0.03cm} q^{2})r^{3}} \right) }{ \alpha \sqrt{ (r^{3} + 8\hspace{0.03cm}\alpha\hspace{0.03cm}q^{2})r^{9}} } 
dr, \label{4D_EGB_LF}
\end{eqnarray}
where $\alpha$, $a$, and $q$ are real parameters (with $\alpha\neq0$), defining a NLED-SF model for the which the metric
\begin{equation}\label{BB_4dEGB}
\boldsymbol{ds^{2}}_{_{B\!B}} = - \left[ 1 \quad\!\!\!\! + \quad\!\!\!\! \frac{r^{2}}{2\alpha}\!\!\left( 1  - \sqrt{  1 + \frac{8\hspace{0.03cm}\alpha\hspace{0.03cm}q^{2} }{r^{3}}  } \quad\!\!\!\!\!\right) \right]\boldsymbol{dt^{2}} + \frac{ \boldsymbol{dr^{2}} }{ \left( 1 - \frac{a^{^{2}}}{r^{^{2}}} \right)\!\!\left[ 1 \quad\!\!\!\! + \quad\!\!\!\! \frac{r^{2}}{2\alpha}\!\!\left( 1  - \sqrt{  1 + \frac{8 \hspace{0.03cm} \alpha \hspace{0.03cm} q^{2} }{r^{3}}  } \quad\!\!\!\!\!\right) \right] }
+ r^{2}\boldsymbol{d\Omega^{2}}, 
\end{equation}
together with the scalar field 
\begin{equation}\label{4GB_BB_SF}
\phi(r) = 2i\tan^{^{\!\!-1}}\!\!\left( \sqrt{\frac{r^{2} - a^{2}}{a^{2}}}\right)    
\end{equation}
is a pure magnetic exact solution of the GR-NLED-SF field equations (\ref{Eqt})$-$(\ref{phi2}), being that the parameter $q$ is the magnetic charge. 

The metric (\ref{4dEGB}) defines  
a black-bounce model of the Simpson-Visser type (\ref{BB_structure}), which smoothly interpolates between the Ellis wormhole metric (if $q=0\neq a$) and a BH metric  
given by 
\begin{equation}\label{4dEGB}
\boldsymbol{ds^{2}}_{_{B\!H}} = - \left[ 1  + \frac{r^{2}}{2\alpha}\!\!\left( 1  - \sqrt{  1 + \frac{8\alpha q^{2} }{r^{3}}  } \quad\!\!\!\!\!\right) \right]\boldsymbol{dt^{2}} + \left[ 1  + \frac{r^{2}}{2\alpha}\!\!\left( 1  - \sqrt{  1 + \frac{8\alpha q^{2} }{r^{3}}  } \quad\!\!\!\!\!\right) \right]^{-1} \boldsymbol{dr^{2}} 
+ r^{2}\boldsymbol{d\Omega^{2}},   
\end{equation}
which was presented in \cite{4DEGB} as a SSS asymptotically Schwarzschild black hole solution of the 4D-Einstein-Gauss-Bonnet (4D-EGB) field equations and therefore it is known as 4D-EGB black hole.

It is worth mentioning that  
the black-bounce model (\ref{BB_4dEGB})  
is defined using the Ellis wormhole as the T-WH metric (\ref{TWH_MT}), and the 4D-EGB black hole  
as the BH metric (\ref{BH_generic}). 

On the other hand, by the mapping $\rho^{2} = r^{2} - a^{2}$, the line element (\ref{BB_4dEGB}) takes the form
\begin{equation}\label{BB_4dEGB_2}
\boldsymbol{ds^{2}}_{_{B\!B}}\!=\! -\!\left[\! 1 \quad\!\!\!\!\! + \quad\!\!\!\!\! \frac{\rho^{2} + a^{2}}{2\alpha}\!\!\left( \!1\!  -\! \sqrt{  1\! +\! \frac{8\alpha q^{2}}{(\rho^{2} \!+\! a^{2})^{\frac{3}{2}}}  } \quad\!\!\!\!\right) \right]\boldsymbol{dt^{2}} \!+\! \left[ 1 \quad\!\!\!\!\!+\quad\!\!\!\!\! \frac{\rho^{2} \!+\! a^{2}}{2\alpha}\!\!\left( 1 \! - \! \sqrt{  1 \!+\! \frac{8\alpha q^{2}}{ (\rho^{2} \!+\! a^{2})^{\frac{3}{2}} }  } \quad\!\!\!\!\right) \right]^{^{\!\!-1}}\!\!\!\!  
\boldsymbol{d\rho^{2}} 
\!+\! \left(\rho^{2}\!+\!a^{2}\right)\boldsymbol{d\Omega^{2}}. 
\end{equation}
This spacetime configuration smoothly interpolates between a black hole geometry [if $a=0\neq q^{2}$ the metric (\ref{BB_4dEGB_2}) reduces to the 4D-EGB black hole]  and  a  traversable  wormhole  of  the  Morris-Thorne  type [if $q^{2}=0\neq a$ the metric (\ref{BB_4dEGB_2}) reduces to the Ellis T-WH].

Hence, adjusting the parameters $q$ and $a$, the spacetime geometry (\ref{BB_4dEGB_2}) $\equiv$ (\ref{BB_4dEGB})   
admits the following interpretations:
\begin{itemize}
%
    \item[i)] The ordinary traversable Ellis wormhole spacetime if ($q=0\neq a$). 
    \item[ii)] A T-WH generalization of Ellis wormhole 
    if ($q\neq0\neq a$ such that $q^{2} + \sqrt{q^{4} -\alpha} < |a|$, which implies absence of horizons)  and  with WH throat located at $\rho_{0}=0$ (of radius $r_{0}=|a|$).
    \item[iii)] An one-way wormhole if ($q^{2} + \sqrt{q^{4} -\alpha} =|a|\neq0$) with a null throat at $\rho_{0}=0$ (of radius $r_{0}=q^{2} + \sqrt{q^{4} -\alpha}=|a|$).
    \item[iv)] The ordinary 4D-EGB black hole spacetime if ($q\neq 0 = a$).   
    \item[v)] A black bounce with a one-way spacelike throat if ($q\neq 0 \neq a$ such that $q^{2} + \sqrt{q^{4} -\alpha} > |a|$). The one-way spacelike throat is localized at $\rho_{0}=0$ (of radius $r_{0}=|a|$), whereas the event horizons are located at $\rho_{h} = \pm \sqrt{ (q^{2} + \sqrt{q^{4} -\alpha})^{2} - a^{2}}\neq0$ (of radius $r_{h}=q^{2} + \sqrt{q^{4} -\alpha}>|a|$).
\end{itemize}
This black-bounce spacetime avoids the Penrose singularity theorem since the scalar field (\ref{4GB_BB_SF}) is imaginary.

\section{Novel type of black-bounce geometry: black bounce without Ellis wormhole as a particular case}\label{BB_non_Ellis}
To date, all the known SSS-AF black bounces \cite{Simpson2019,Simpson_2,Simpson_3,
Charged_rotating_Bb,Simpson_4} 
are of the form (\ref{BB_structure}) $\equiv$ (\ref{generalBB_BBC_SV}), and hence all they 
have the Ellis wormhole as a particular case, i.e. they are of the Simpson-Visser type, and have the form 
\begin{equation}\label{Fuera_SV}
\boldsymbol{ds^{2}} = - n(\rho)\boldsymbol{dt^{2}} + \frac{\boldsymbol{dr^{2}}}{m(\rho)}  + (\rho^{2} + a^{2})\boldsymbol{d\Omega^{2}} 
\quad\quad\quad \textup{with}\quad\quad\quad n(\rho)=m(\rho).
\end{equation}
In virtue of black-bounce method (Sec. \ref{method}), in the following we will show for the first time a novel type of black bounce metric that cannot be reduced to the Ellis wormhole solution, i.e. we will obtain a black-bounce solution of the form (\ref{Fuera_SV}) with $n(\rho) \neq m(\rho)$. \\ 

{\bf Novel type of black-bounce solution:} 
The following NLED-SF theory defined by a scalar potential and a NLED model, given respectively by
\begin{eqnarray}
\mathscr{U}(\phi) &=& -\frac{1}{48}\left( 1 + \frac{\phi^{2}}{4} \right)^{\!\!\!^{3}} \left[ \frac{8\beta_{_{0}}}{3}\left( \phi^{2} - \frac{44}{7} \right)\sqrt{ \phi^{2} + 4 } - \tilde{\beta}_{_{0}} 
\left(3\phi^{2} - 4\right)  \right], \label{non_ELLis_model_U}\\  
\mathcal{L}(\mathcal{F}) &=& - \frac{1}{2} \mathcal{F} + \frac{2|q|}{3}|\mathcal{F}|^{^{\frac{3}{2}}} - \frac{q^{2}}{8}\mathcal{F}^{^{2}} + \sigma_{_{0}}\!\!\left( |\mathcal{F}|^{^{\frac{5}{4}}} - \frac{11|q|}{14}|\mathcal{F}|^{^{\frac{7}{4}}}   + \frac{q^{2}}{9}|\mathcal{F}|^{^{\frac{9}{4}}}\right), \label{NLmax1}
\end{eqnarray}
where  $\tilde{\beta}_{_{0}}$, $\beta_{_{0}}$, and $\sigma_{_{0}}$  are real parameters, admits the following metric:
\begin{equation}\label{non_ELLis_WH}
\boldsymbol{ds^{2}}_{_{W\!H}} = - e^{\!^{ -\frac{q^{2}}{r^{2}} } }\boldsymbol{dt^{2}} + \frac{\boldsymbol{dr^{2}}}{1 - \frac{q^{2}}{r^{2}} } + r^{2}\boldsymbol{d\Omega^{2}} 
\end{equation}
 for the value of the parameters $\tilde{\beta}_{_{0}} = 1 /q^{2}$, $\beta_{_{0}}=0=\sigma_{_{0}}$, together with the scalar field
\begin{equation}\label{SF_non_Ellis}
\phi(r) = 2i \sqrt{ 1 - \frac{q^{2}}{r^{2}} } %
\end{equation} %
as a pure magnetic exact solution of the GR-NLED-SF field equations (\ref{Eqt})$-$(\ref{phi2}). 
The metric (\ref{non_ELLis_WH}) is a nontrivial redshift function modification of the T-WH metric (\ref{EllisWH}) and was recently derived in \cite{Pe_Fe}, 
whereas, for the case $\tilde{\beta}_{_{0}} = 1 /q^{2}$,   
$\beta_{_{0}}= m/|q|^{3}$, and  
$\sigma_{_{0}}=2m/\sqrt{|q|}$, 
it defines a NLED-SF model for the which the metric
\begin{equation}\label{non_ELLis_BB}
\boldsymbol{ds^{2}}_{_{B\!B}} = - \left( 1 - \frac{2m}{r} \right)e^{\!^{ -\frac{q^{2}}{r^{2}} } }\boldsymbol{dt^{2}} + \frac{\boldsymbol{dr^{2}}}{\big(1 - \frac{q^{2}}{r^{2}} \big)\big(1 - \frac{2m}{r} \big)} + r^{2}\boldsymbol{d\Omega^{2}} \end{equation}
together with the scalar field (\ref{SF_non_Ellis}), is a pure magnetic exact solution of the GR-NLED-SF field equations (\ref{Eqt})$-$(\ref{phi2}).  
The metric (\ref{non_ELLis_BB}) has a black-bounce structure of the type (\ref{generalBB}) with nontrivial $\Psi(r)$ function, and  smoothly interpolates between the Schwarzschild black hole (if $q=0\neq m$) and a traversable wormhole (if $q\neq0= m$) with metric (\ref{non_ELLis_WH}). 
In terms of the radial $\rho$-coordinate, $r = \sqrt{\rho^{2} + q^{2}} \geq |q|$, 
the metric (\ref{non_ELLis_BB}) takes the form
\begin{equation}\label{non_ELLis_BB_rho}
\boldsymbol{ds^{2}}_{_{B\!B}} = - \left( 1 - \frac{2m}{\sqrt{\rho^{2} + q^{2}}} \right)e^{\!^{ -\frac{q^{2}}{\rho^{2} + q^{2}} } }\boldsymbol{dt^{2}} + \frac{\boldsymbol{d\rho^{2}}}{\left( 1 - \frac{2m}{\sqrt{\rho^{2} + q^{2}}} \right)} + \left(\rho^{2}+q^{2}\right)\boldsymbol{d\Omega^{2}}; 
\end{equation}
since this metric yields $|g_{_{tt}}g_{_{\rho\rho}}| = e^{\!^{ -\frac{q^{2}}{\rho^{2} + q^{2}} } }$, it is not of  Simpson-Visser black bounce (\ref{B_bounce}) type because we must have $|g_{_{tt}}g_{_{\rho\rho}}| =1$ for all  
the SSS black bounces of the Simpson-Visser type. Hence, the metric  (\ref{non_ELLis_BB}) $\equiv$ (\ref{non_ELLis_BB_rho}) defines a new type of black bounce. 
Adjusting the parameters $m$ and $q$, the spacetime configuration (\ref{non_ELLis_BB_rho}) $\equiv$ (\ref{non_ELLis_BB}) represents either:
%
\begin{itemize}
%
    \item[i)] The traversable wormhole metric (\ref{non_ELLis_WH}) if ($q\neq 0=m $). 
    \item[ii)] A T-WH generalization of (\ref{non_ELLis_WH}) if ($q\neq0\neq m$ such that $ |q| > 2m$) with T-WH throat at $\rho_{0} =0$ (of radius $r_{0}=|q|$).
    \item[iii)] An one-way wormhole if ($|q|=2m\neq0$) with a null throat at $\rho_{0} =0$ (of radius $r_{0}=|q|$).
    \item[iv)] The ordinary Schwarzschild black hole spacetime if ($q =0\neq m$).   
    \item[v)] A black bounce if ($q\neq 0 \neq m$ such that $|q| < 2m$), with an one-way spacelike throat at $\rho_{0} =0$  (of radius $r_{0}=|q|$), and with event horizons at $\rho_{h} = \pm\sqrt{4m^{2} - q^{2} }\neq0$  (of radius $r_{h}=2m>|q|$).
\end{itemize}
%
Finally, it is important to emphasize that the NLED Lagrangian density (\ref{NLmax1})  
needed to generate the new black bounce solution (\ref{non_ELLis_BB}) $\equiv$ (\ref{non_ELLis_BB_rho}) 
reduces to Maxwell theory in the limit of weak field. I.e., $\mathcal{L} \rightarrow \kappa \mathcal{F} $ and $\mathcal{L}_{\mathcal{F}} \rightarrow \kappa $ (being that $\kappa$ is a constant) when $\mathcal{F}\rightarrow0$, in contrast to $\mathcal{L}(\mathcal{F})$ functions (\ref{U_L_SV}), (\ref{NLED_SF_CABB}), and (\ref{4D_EGB_LF}) which do not satisfy this important physical NLED condition. %
However, because the spacetime metric (\ref{non_ELLis_BB})  
requires the phantom scalar fields (\ref{SF_non_Ellis})  
in order to support its T-WH structures, yield that these spacetimes violate the WEC.

\section{Conclusion}

In this work the construction of several 
black-bounce geometries as exact solutions of the  Einstein-nonlinear electrodynamics gravity coupled to a phantom scalar field have been presented. Specifically, we show three black bounces of the Simpson-Visser type:  
the Simpson-Visser, the canonical acoustic, and the 4D-EGB, all of them as
purely magnetic, exact solutions of GR-NLED-SF. 
Moreover, we obtained a novel black-bounce solution which, in contrast of the Simpson-Visser type, does not have the Ellis  wormhole solution as a particular case. Therefore, a new type of black bounce of the form 
\begin{equation}\label{Canate}
\boldsymbol{ds^{2}} = - n(\rho)\boldsymbol{dt^{2}} + \frac{\boldsymbol{dr^{2}}}{m(\rho)}  + (\rho^{2} + a^{2})\boldsymbol{d\Omega^{2}}, 
\end{equation}
with $n(\rho)\neq m(\rho)$ is established.  
Particularly, the nonlinear electrodynamics model needed to generate this novel solution, which, in the limit of weak field, becomes the Maxwell field.  
In contrast with the Simpson-Visser black bounces type, i.e. 
$n(\rho)= m(\rho)$ in all cases, and for which the Maxwell limit is not achieved in the cases we analyzed (\ref{U_L_SV}), (\ref{NLED_SF_CABB}), and (\ref{4D_EGB_LF}).
However, all the presented solutions (\ref{BH_WH_2}), (\ref{BH_WH_CA}), (\ref{BB_4dEGB_2}), and (\ref{non_ELLis_BB_rho}), Simpson-Visser type or not, require a phantom scalar field as a source in order 
to support the traversable wormhole structure.   
The existence of this phantom scalar field implies that the black-bounce spacetime violates the weak energy condition, 
and therefore the Penrose singularity theorem is avoided.\\
While this paper was being written, we learned of the
results from a similar work \cite{Bronnikov_BBs}. There  
the field sources for Simpson-Visser spacetimes are presented; however, in their approach only  black bounce of the Simpson–Visser type, i.e. of the form (\ref{Canate}) with $n(\rho)=m(\rho)$ is studied (See Eq. (1) of the Ref. \cite{Bronnikov_BBs}). Hence, black bounces without Ellis wormhole as a particular case, e.g. our solution (\ref{non_ELLis_BB_rho}), cannot be derived by the approach established in \cite{Bronnikov_BBs}. \\  

\textbf{Acknowledgments.} We thank Kirill Bronnikov for very useful discussions. This work was partially supported by Coordena\c c\~ao de Aperfei\c coamento de Pessoal de N\'ivel Superior-Brasil (CAPES) - C\'odigo de Financiamento 001.

\appendix 
\section{\bf Curvature invariants $R$, $R_{\alpha\beta}R^{\alpha\beta}$ and $R_{\alpha\beta\mu\nu}R^{\alpha\beta\mu\nu}$  in terms of the Schwarzschild coordinates }

For the metric (\ref{generalBB}) the curvature invariants $R$, $R_{\alpha\beta}R^{\alpha\beta}$ and $R_{\alpha\beta\mu\nu}R^{\alpha\beta\mu\nu}$ are given by 
\begin{eqnarray}
R &=&\frac{1}{2r^{2}}\bigg\{\!\!\left( r - b \right)\!\!\left[4\mathcal{M}'' - \left( r - 2\quad\!\!\!\!\!\mathcal{M}\right)\!\!\left( 2\Psi'' + \Psi'^{2}\right)\!\right] \!+\! \left[ 6\!\left(r-b\right)\!\mathcal{M}' \!+\! \left(r-2\mathcal{M}\right)\!b' - 4r + 2\mathcal{M} +  3\quad\!\!\!\!\!b \right]\Psi' \nonumber\\
&& \quad\!\!\!\! + \quad\!\!\!\!  \left(8 - \frac{6\quad\!\!\!\!\!b}{r} - 2\quad\!\!\!\!\!b'\right)\!\mathcal{M}' + \left(4 - \frac{6\quad\!\!\!\!\!\mathcal{M}}{r}\right)\!b' + \frac{6\quad\!\!\!\!\!\mathcal{M}\quad\!\!\!\!\!b}{r^{2}} \bigg\},  
\label{Ext_R_BB_gen} \\
\nonumber\\
R_{\alpha\beta}R^{\alpha\beta}\!&=&\!\frac{1}{4r^{4}}\bigg\{\!\left( r \!-\! b \right)\!\left( r \!-\! 2\mathcal{M} \right)\!\left(\!\Psi'' \!+\! \frac{\Psi'^{2}}{2}\right) \!-\!2(r\!-\!b)\mathcal{M}'' \!-\! \left[3(r\!-\!b)\mathcal{M}' \!+\! \frac{1}{2}(r\!-\!2\mathcal{M})b' - 2r + \mathcal{M} + \frac{3}{2}b\right]\!\!\Psi'  \nonumber\\
 &&\quad\!\!\!\!\! + \quad\!\!\!\!\!\left(\!\frac{b}{r}\!-\!b'\!\right)\!\!\left(\!\frac{\mathcal{M}}{r}\!-\! \mathcal{M}'\!\right)\!\!\bigg\}^{\!\!2} \!+\! \frac{1}{4r^{4}}\bigg\{\!2(b-r)\mathcal{M}'' \!+\! (r-2\mathcal{M})(r-b)\!\!\left( \Psi'' + \frac{\Psi'^{2}}{2} \right) - \frac{1}{2}\bigg[ (r-2\mathcal{M})b'     \nonumber\\
 && \quad\!\!\!\!\! +\quad\!\!\!\!\! 6(r-b)\mathcal{M}' - \!\left(1-\frac{8\mathcal{M}}{r}\right)\!b - 6\mathcal{M}\!\bigg]\!\Psi' + \left(\!b' - \frac{b}{r}\!\right)\!\!\!\left(\mathcal{M}' + \frac{3\mathcal{M}}{r} - 2\right)\!
 \bigg\}^{\!\!2} + \frac{1}{2r^{4}}\!\bigg\{ \!\!\left(1 - \frac{b}{r}\right)\!\!\!\bigg[4\mathcal{M}' \nonumber\\  
 && 
 \quad\!\!\!\!\! - (r-2\mathcal{M}) \Psi' \bigg] + \left( 1 - \frac{2\mathcal{M}}{r}\right)\!b'  +  \left( 1 + \frac{2\mathcal{M}}{r}\right)\!\!\frac{b}{r} \bigg\}^{\!\!2},
\label{Ext_RR2_BB_gen} \\
\nonumber\\
R_{\alpha\beta\mu\nu}R^{\alpha\beta\mu\nu}\!&=&\!\frac{1}{r^{8}}\bigg\{ r^{2}(r-b)(r-2\mathcal{M})\left( \Psi'' + \frac{\Psi'^{2}}{2}\right) - 2r^{2}(r-b)\mathcal{M}'' - 3r\bigg[ r(r-b)\mathcal{M}' + \frac{r(r-2\mathcal{M})b'}{6} \nonumber\\
&&\quad\!\!\!\!\! - \quad\!\!\!\!\!\!\!\left(\!r\!-\!\frac{4b}{3}\!\right)\!\mathcal{M} \!-\! \frac{rb}{6} \bigg]\!\Psi' \!+\! \left(4r\!-\!5b\!+\!rb'\right)\left(r\mathcal{M}' \!\!-\! \mathcal{M}\right)\!\bigg\}^{\!\!2} \!\!+\! \frac{2(b\! -\! r)^{^{\!2}}\!\left[ r\left(2\mathcal{M} \!- r\right)\Psi' \!+\! 2r\mathcal{M} \!-\! 2\mathcal{M}\right]^{^{\!2}}}{r^{8}}  \nonumber\\
&&\quad\!\!\!\! + \quad\!\!\!\! \frac{2}{r^{8}}\left[ r(2\mathcal{M} - r)b' +2r(b-r)\mathcal{M}' +2(r - 2b)\mathcal{M} + r b \right]^{2}
+ \frac{4\left( 2\mathcal{M}b - 2r\mathcal{M} - r b  \right)^{2}}{r^{8}}.
\label{Ext_RR4_BB_gen}
\end{eqnarray}
\section{\bf Field equations}
%
In this appendix we include the explicit form of the field equations that are satisfied by the metric ansatz (\ref{BB_structure}). 

\begin{eqnarray}
&&
G_{t}{}^{t} = \frac{a^{2}}{r^{4}}\left( 1 - \frac{4\mathcal{M} }{r} + 2\mathcal{M}' \right) - \frac{2\mathcal{M}'}{r^{2}}, \quad\quad\quad\quad\quad G_{r}{}^{r} = \frac{a^{2}}{r^{4}}\left( 2\mathcal{M}' - 1 \right) - \frac{2\mathcal{M}'}{r^{2}}, \\
&&
G_{\theta}{}^{\theta}=G_{\phi}{}^{\phi}= \frac{a^{2}}{r^{3}}\left( \mathcal{M}'' - \frac{\mathcal{M}' }{r} - \frac{\mathcal{M} }{r^{2}} + \frac{1}{r} \right) - \frac{\mathcal{M}'' }{r}.
\end{eqnarray}
The nontrivial components of the energy-momentum tensor of self-interacting scalar field are
\begin{eqnarray}
&&8\pi (E_{t}{}^{t})\!_{_{_{S \! F }}} = 8\pi (E_{\theta}{}^{\theta})\!_{_{_{S \! F }}} = 8\pi (E_{\varphi}{}^{\varphi})\!_{_{_{S \! F }}} = - \frac{1}{4} \!\left(1-\frac{a^{2}}{r^{2}}\right)\!\!\left(1-\frac{2\mathcal{M}}{r}\right) \phi'^{2} - \mathscr{U}, \label{Ett} \\ 
&&8\pi (E_{r}{}^{r})\!_{_{_{S \! F }}} = \frac{1}{4} \left(1-\frac{a^{2}}{r^{2}}\right)\!\!\left(1-\frac{2\mathcal{M}}{r}\right) \phi'^{2} - \mathscr{U}. \label{Err}
\end{eqnarray}
The  energy-momentum tensor components  for NLED,  assuming the SSS spacetime with metric (\ref{BB_structure}), the purely magnetic field  (\ref{magnetica}), and a generic Lagrangian density $\mathcal{L}(F)$ are given by 
\begin{eqnarray}
8\pi (E_{t}{}^{t})\!_{_{_{N \! L \! E \! D}}} = 8\pi (E_{r}{}^{r})\!_{_{_{N \! L \! E \! D}}} =  -2\mathcal{L}, 
\quad\quad\quad 8\pi (E_{\theta}{}^{\theta})\!_{_{_{N \! L \! E \! D}}} = 8\pi (E_{\varphi}{}^{\varphi})\!_{_{_{N \! L \! E \! D}}} =  2(2\mathcal{F}\mathcal{L}_{\mathcal{F}} - \mathcal{L}). 
\end{eqnarray}


Inserting the above given components in the field equations written as
$C_{\alpha}{}^{\beta} = G_{\alpha}{}^{\beta} - 8\pi [ (E_{\alpha}{}^{\beta})\!_{_{_{S \! F }}} + (E_{\alpha}{}^{\beta})\!_{_{_{N \! L \! E \! D}}} ] = 0$, we obtain that the GR-NLED-SF field equations for the metric ansatz (\ref{BB_structure}) and the magnetic field (\ref{magnetica}) take the form:
\begin{eqnarray}
&&\!C_{t}{}^{t}\!=\!0\hspace{0.4cm}\!\Rightarrow\!\hspace{0.4cm}
\frac{a^{2}}{r^{4}}\!\left(\!1 - \frac{4\mathcal{M}}{r} + 2\mathcal{M}'\!\right) - \frac{2\mathcal{M}'}{r^{2}} + \frac{1}{4} \!\left(1-\frac{a^{2}}{r^{2}}\right)\!\!\left(\!1-\frac{2\mathcal{M}}{r}\!\right)\! \phi'^{2} + \mathscr{U} + 2\mathcal{L} =0, %
\\
&&\nonumber\\
&&\!C_{r}{}^{r}\!=\!0\hspace{0.4cm}\!\Rightarrow\!\hspace{0.4cm} \frac{a^{2}}{r^{4}}\!\left( 2\mathcal{M}' - 1 \right) - \frac{2\mathcal{M}' }{r^{2}} 
- \frac{1}{ 4 }\!\left(1-\frac{a^{2}}{r^{2}}\right)\!\!\left(\!1-\frac{2\mathcal{M}}{r}\!\right)\! \phi'^{2} + \mathscr{U} %
+ 2\mathcal{L} = 0, %
\\
&&\nonumber\\
&&\!C_{\theta}{}^{\theta}\!=\!0\hspace{0.4cm}\!\Rightarrow\!\hspace{0.4cm} \frac{a^{2}}{r^{3}}\!\!\left(\!\mathcal{M}'' \!-\! \frac{\mathcal{M}'}{r} \!-\! \frac{\mathcal{M}}{r^{2}} \!+\! \frac{1}{r} \!\right) - \frac{\mathcal{M}''}{r} + \frac{1}{4} \!\left(\!1\!-\!\frac{a^{2}}{r^{2}}\!\right)\!\!\left(\!1\!-\!\frac{2\mathcal{M}}{r}\!\right)\! \phi'^{2} + \mathscr{U} - 2(2\mathcal{F}\mathcal{L}_{\mathcal{F}} - \mathcal{L})  =0, %
\end{eqnarray}
whereas the scalar field should satisfy
\begin{equation}%
2r\phi'' + \left( 4 + \frac{4\mathcal{M} - 4r\mathcal{M}'}{r - 2\mathcal{M}} + \frac{ 2a^{2} }{ r^{2} - a^{2}} \right)\phi' -  \frac{4r^{4}}{(r^{2} - a^{2})(r - 2\mathcal{M})}  \dot{\mathscr{U}}  = 0. 
\end{equation}


\section{\bf Avoiding the black hole singularity problem}\label{Avoiding_singularity}

{\bf Avoiding the black hole singularity problem in general relativity:}
The Penrose singularity theorem \cite{Penrose}, and its modern variants and extensions (see for instance \cite{Senovilla}), demonstrates that in a gravitational collapse with the assumption that the Einstein field equations hold, 
once a closed trapped surface $\mathcal{S}$ is formed, which describes the inner region of an black hole event horizon, then in some region of spacetime contained in the causal future $\mathcal{J}^{+}(\mathcal{S})$ of $\mathcal{S}$, at least 
one of the following must hold in order to avoid a spacetime singularity: 
\begin{itemize}
\item The weak energy condition (WEC) is violated.
\item Global hyperbolicity breaks down.
\end{itemize}

{\bf Null and weak energy conditions in GR:}
For a energy-momentum tensor $T_{\mu\nu}$, the null energy condition (NEC) stipulates that for every null vector, $n^{\alpha}$ yields $T_{\mu\nu}n^{\mu}n^{\nu}\geq0$.
Following \cite{WEC}, for a diagonal energy-momentum tensor $(T_{\alpha\beta})=diag \left( T_{tt},T_{rr},T_{\theta\theta},T_{\varphi\varphi} \right)$, which can be
conveniently written as 
\begin{equation}\label{diagonalEab}
T_{\alpha}{}^{\beta} = - \rho_{t} \hskip.05cm \delta_{\alpha}{}^{t}\delta_{t}{}^{\beta} + P_{r} \hskip.05cm \delta_{\alpha}{}^{r}\delta_{r}{}^{\beta} + P_{\theta} \hskip.05cm \delta_{\alpha}{}^{\theta}\delta_{\theta}{}^{\beta} + P_{\varphi} \hskip.05cm \delta_{\alpha}{}^{\varphi}\delta_{\varphi}{}^{\beta} ,
\end{equation}
 where $\rho_{t}$ may be interpreted as the rest energy density of the matter,  
 whereas $P_{r}$, $P_{\theta}$, and $P_{\varphi}$ are respectively the pressures along the $r$, $\theta$, and $\varphi$ directions. In terms of (\ref{diagonalEab}) the NEC implies
\begin{equation}\label{NEC_Eab}
\rho_{t} + P_{a} \geq0 \quad\quad\textup{with}\quad\quad  a = \{ r, \theta, \varphi \}.
\end{equation}
The weak energy condition (WEC) states that for any timelike vector $\boldsymbol{k} = k^{\mu}\partial_{\mu}$ (i.e., $k_{\mu}k^{\mu}<0$), the energy-momentum tensor obeys the inequality
$T_{\mu\nu}k^{\mu}k^{\nu} \geq 0$, which means that the local energy density $\rho_{\!_{_{loc}}}= T_{\mu\nu}k^{\mu}k^{\nu}$ as measured by any observer with timelike vector 
$\boldsymbol{k}$ is a non-negative quantity.  
For an energy-momentum tensor of the form (\ref{diagonalEab}), the WEC will be satisfied if and only if
\begin{equation}\label{WEC}
\rho_{t} = - T_{t}{}^{t} \geq0, \quad\quad\quad\quad \rho_{t} + P_{a} \geq0 \quad\textup{with}\quad  a = \{ r, \theta, \varphi \}.
\end{equation}

\begin{itemize}
\item {\bf WEC for a self-interacting scalar field $(E_{\alpha}{}^{\beta})\!_{_{_{S \! F }}}$}

Now, by  using (\ref{E_SF}), (\ref{Ett}), and (\ref{Err}) yield
\begin{eqnarray}
&& 8\pi (\rho_{t})\!_{_{_{S \! F }}} = - 8\pi (P_{\theta})\!_{_{_{S \! F }}} = - 8\pi (P_{\varphi})\!_{_{_{S \! F }}} = \frac{1}{4} \!\left(1-\frac{a^{2}}{r^{2}}\right)\!\!\left(1-\frac{2\mathcal{M}}{r}\right) \phi'^{2} + \mathscr{U}, \\ 
&& 8\pi (P_{r})\!_{_{_{S \! F }}} = \frac{1}{ 4 } \left(1-\frac{a^{2}}{r^{2}}\right)\!\!\left(1-\frac{2\mathcal{M}}{r}\right) \phi'^{2} - \mathscr{U},   
\end{eqnarray}
since $(\rho_{t})\!_{_{_{S \! F }}} + (P_{a})\!_{_{_{S \! F }}}=0$ for all $a=\theta$, $\varphi$, the tensor $(E_{\alpha}{}^{\beta})\!_{_{_{S \! F }}}$ satisfies the WEC if 
\begin{equation}\label{WEC_SF}
8\pi (\rho_{t})\!_{_{_{S \! F }}} = \frac{1}{4} \!\left(1-\frac{a^{2}}{r^{2}}\right)\!\!\left(1-\frac{2\mathcal{M}}{r}\right) \phi'^{2} + \mathscr{U} \geq0, \quad 8\pi(\rho_{t})\!_{_{_{S \! F }}} + 8\pi (P_{r})\!_{_{_{S \! F }}}  = \frac{1}{2} \!\left(1-\frac{a^{2}}{r^{2}}\right)\!\!\left(1-\frac{2\mathcal{M}}{r}\right) \phi'^{2}\geq0.
\end{equation}
\item {\bf WEC for the nonlinear electromagnetic field $(E_{\alpha}{}^{\beta})\!_{_{_{N \! L \! E \! D}}}$}

By using (\ref{NLED_EM}) and (\ref{E_nled}),
%
\begin{eqnarray}%
8\pi (\rho_{t})\!_{_{_{N \! L \! E \! D}}} = - 8\pi (P_{r})\!_{_{_{N \! L \! E \! D}}} =  2\mathcal{L},   
\quad\quad\quad 8\pi (P_{\theta})\!_{_{_{N \! L \! E \! D}}} = 8\pi (P_{\varphi})\!_{_{_{N \! L \! E \! D}}} =  2(2\mathcal{F}\mathcal{L}_{\mathcal{F}} - \mathcal{L}), 
\end{eqnarray}
since $\rho_{t} + P_{r}=0$, the tensor $(E_{\alpha}{}^{\beta})\!_{_{_{N \! L \! E \! D}}}$ satisfies the WEC if
\begin{equation}\label{WEC_NLED}
8\pi (\rho_{t})\!_{_{_{N \! L \! E \! D}}} =  2\mathcal{L}\geq0, \quad\quad  8\pi(\rho_{t})\!_{_{_{N \! L \! E \! D}}} + 8\pi (\rho_{\theta})\!_{_{_{N \! L \! E \! D}}}   = 8\pi(\rho_{t})\!_{_{_{N \! L \! E \! D}}} + 8\pi(\rho_{\varphi})\!_{_{_{N \! L \! E \! D}}}  =  4\mathcal{F}\mathcal{L}_{\mathcal{F}}\geq0.
\end{equation}
\item {\bf  WEC for the effective energy-momentum tensor $(E_{\alpha}{}^{\beta})\!_{_{_{e\!f\!f\!}}}=(E_{\alpha}{}^{\beta})\!_{_{_{S \! F }}}+ (E_{\alpha}{}^{\beta})\!_{_{_{N \! L \! E \! D}}}$}
\begin{eqnarray}
&&8\pi (\rho_{t})\!_{_{_{e\!f\!f\!}}} = \frac{1}{4} \!\left(1-\frac{a^{2}}{r^{2}}\right)\!\!\left(1-\frac{2\mathcal{M}}{r}\right) \phi'^{2} + \mathscr{U} + 2\mathcal{L} \\
&&8\pi (P_{r})\!_{_{_{e\!f\!f\!}}} = \frac{1}{ 4 } \left(1-\frac{a^{2}}{r^{2}}\right)\!\!\left(1-\frac{2\mathcal{M}}{r}\right) \phi'^{2} - \mathscr{U} - 2\mathcal{L} \\
&&8\pi (P_{\theta})\!_{_{_{e\!f\!f\!}}} = 8\pi (P_{\varphi})\!_{_{_{e\!f\!f\!}}} = -\frac{1}{4} \!\left(1-\frac{a^{2}}{r^{2}}\right)\!\!\left(1-\frac{2\mathcal{M}}{r}\right) \phi'^{2} - \mathscr{U} + 2(2\mathcal{F}\mathcal{L}_{\mathcal{F}} - \mathcal{L}).
\end{eqnarray}
So, the tensor $(E_{\alpha}{}^{\beta})\!_{_{_{e\!f\!f\!}}}$ satisfies the WEC if
\begin{eqnarray}
&&8\pi (\rho_{t})\!_{_{_{e\!f\!f\!}}} = \frac{1}{4} \!\left(1-\frac{a^{2}}{r^{2}}\right)\!\!\left(1-\frac{2\mathcal{M}}{r}\right) \phi'^{2} + \mathscr{U} + 2\mathcal{L} \geq 0, \\
\nonumber\\
&& 8\pi (\rho_{t})\!_{_{_{e\!f\!f\!}}} + 8\pi (P_{r})\!_{_{_{e\!f\!f\!}}} = \frac{1}{4} \!\left(1-\frac{a^{2}}{r^{2}}\right)\!\!\left(1-\frac{2\mathcal{M}}{r}\right) \phi'^{2} \geq 0, \label{WEC_effect}\\
\nonumber\\
&& 8\pi (\rho_{t})\!_{_{_{e\!f\!f\!}}} + 8\pi (P_{\theta})\!_{_{_{e\!f\!f\!}}} = 8\pi (\rho_{t})\!_{_{_{e\!f\!f\!}}} + 8\pi (P_{\varphi})\!_{_{_{e\!f\!f\!}}} = 4\mathcal{F}\mathcal{L}_{\mathcal{F}}\geq0.
\end{eqnarray}
\end{itemize}


\section*{Bibliography}
\begin{thebibliography}{99}

\bibitem{Simpson2019} A. Simpson and M. Visser, {\it Black-bounce to traversable wormhole},
JCAP {\bf 02} (2019) 042. 



\bibitem{rotating_Bb} J. Mazza, E. Franzin and S. Liberati, {\it A novel family of rotating black hole mimickers}, JCAP {\bf04} (2021) 082.

\bibitem{Charged_rotating_Bb} E. Franzin, S. Liberati, Jacopo Mazza, A. Simpson,  M. Visser, {\it Charged black-bounce spacetimes}, JCAP {\bf07} (2021) 036.

\bibitem{Simpson_4} F.S.N. Lobo, M.E. Rodrigues, M.V.d.S. Silva, A. Simpson and M. Visser, {\it Novel black-bounce spacetimes: Wormholes, regularity, energy conditions, and causal structure}, 
Phys. Rev. D {\bf 103}, 084052 (2021). 




\bibitem{Bertotti} B. Bertotti, L. Iess, P. Tortora, {\it A test of general relativity using radio links with the Cassini spacecraft}, Nature {\bf 425} 374 (2003).

\bibitem{Williams} J. G. Williams,  S. G. Turyshev, D. H. Boggs, Phys. Rev. Lett. {\bf93} 261101 (2004).

\bibitem{Will1} C. M. Will, {\it Theory and Experiment in Gravitational Physics} (Cambridge: Cambridge University
Press) (1993).

\bibitem{Will2}  C. M. Will, {\it The Confrontation between General Relativity and Experiment},  Living Rev. Relativ. {\bf9} 3 (2006).

\bibitem{Psaltis}  D. Psaltis,  {\it Probes and Tests of Strong-Field Gravity with Observations in the Electromagnetic Spectrum}, Living Rev. Relativ. {\bf11} 9 (2008).


\bibitem{Hawking1976}  S. W. Hawking, {\it Breakdown of predictability in gravitational collapse}. Phys. Rev. D. {\bf14} 2460–2473 (1976); 

\bibitem{Carroll2001} S. M. Carroll, {\it The Cosmological Constant }, Living. Rev. Rel. {\bf4} 1 (2001).

\bibitem{Romero2013} Gustavo E. Romero, {\it Adversus Singularitates: The Ontology of Space–Time Singularities},
Foundations of Science 18 (2):297-306 (2013).



\bibitem{Narlikar1985} J. V. Narlikar, T. Padmanabhan, {\it Creation-field cosmology: A possible solution to singularity, horizon, and flatness problems}, Phys. Rev. D {\bf32} 1928 (1985).

\bibitem{Rendall1996} A. D. Rendall, {\it The initial singularity in solutions of the Einstein–Vlasov system of Bianchi type I}, J. Math. Phys. {\bf37} 438 (1996).

\bibitem{Hayward2006}S.A. Hayward, {\it Formation and evaporation of non-singular black holes}, Phys. Rev. Lett. {\bf96} (3), 031103 (2006).

\bibitem{Ashtekar2009} A. Ashtekar, {\it Singularity resolution in loop quantum cosmology:
a brief overview}, Journal of Physics: Conference Series {\bf189} (2009) 012003. 

\bibitem{Kuntz2020} I. Kuntz, R. Casadio, {\it Singularity avoidance in quantum gravity}, Phys. Lett., B {\bf802} 135219 (2020).


\bibitem{Bardeen} J. Bardeen, {\it Non singular general relativistic gravitational collapse}, Proc. GR5, Tiblisi, USSR (1968).

\bibitem{quantum_BH} P. H\'aj\'i\v{c}ek and C. Kiefer, {\it Singularity avoidance by collapsing shells in quantum gravity}, Int. J. Mod. Phys. D {\bf10} 775 (2001); C. Kiefer, {\it Quantum black hole without singularity}, arXiv:1512.08346 [gr-qc] (2015); C. Kiefer, {\it Aspects of Quantum Black Holes
}, J. Phys.: Conf. Ser. {\bf1612} 012017 (2020); R. V. Maluf and J. C. S. Neves, {\it Bardeen regular black hole as a quantum-corrected Schwarzschild black hole}, Int. J. Mod. Phys. D {\bf28}, 1950048 (2019).


\bibitem{Ayon_Garcia} E. Ay\'on–Beato and A. Garc\'ia, {\it Regular Black Hole in General Relativity Coupled to Nonlinear Electrodynamics}, Phys. Rev. Lett. {\bf80}, 5056 (1998); {\it Non-Singular Charged Black Hole Solution for Non-Linear Source}, Gen. Rel. Gravit., {\bf31}, 629 (1999); {\it New Regular Black Hole Solution from Nonlinear Electrodynamics}, Phys. Lett., B {\bf464}, 25 (1999); {\it Four Parametric Regular Black Hole Solution}, Gen. Rel. Grav. {\bf37}, 635 (2005).

\bibitem{Ayon_Garcia2000} E. Ay\'on–Beato and A. Garc\'ia, {\it The Bardeen Model as a Nonlinear Magnetic Monopole}, Phys. Lett. B {\bf493}, 149 (2000).


\bibitem{Penrose}R. Penrose, {\it Gravitational Collapse and Space-Time Singularities}, Phys. Rev. Lett. {\bf 14}, 57 (1965).

\bibitem{Penrose_GC} R. Penrose, {\it Gravitational Collapse: The Role of General Relativity},  Nuovo Cimento. Rivista Serie. {\bf1}: 252-276 (1969), [Gen. Rel. Grav. 34, 1141 (2002)]; R. M. Wald, {\it Gravitational Collapse and Cosmic Censorship}, arXiv:9710068 [gr-qc] (1997).

\bibitem{Wald2} R. M. Wald, {\it Final States of Gravitational Collapse}, Phys. Rev. Lett. {\bf 26}, 1653 (1971).

\bibitem{Senovilla} J.M.M. Senovilla, D. Garfinkle, {\it The 1965 Penrose singularity theorem}, Class. Quant. Grav. {\bf32}, 124008 (2015).

\bibitem{BronnikovRBH2000} \textcolor{black}{ K. A. Bronnikov, {\it Comment on “Regular Black Hole in General Relativity Coupled to Nonlinear Electrodynamics}, Phys. Rev. Lett. {\bf85}, 4641 (2000).}

\bibitem{BronnikovRBH2001} \textcolor{black}{ K. A. Bronnikov, {\it Regular Magnetic Black Holes and Monopoles from Nonlinear Electrodynamics}, Phys.Rev.D {\bf63}, 044005 (2001).}

\bibitem{Durrer2013} R. Durrer, A. Neronov, {\it Cosmological Magnetic Fields: Their Generation, Evolution and Observation}, Astron. Astrophys. Rev. {\bf 21} 62 (2013).

\bibitem{Kunze2013} K.E. Kunze, {\it Cosmological magnetic fields}, Plasma Phys. Control. Fusion {\bf55} 124026 (2013).



\bibitem{Camara2004} C. S. Camara, M. R. de Garcia Maia, J. C. Carvalho, J. A. S. Lima, {\it Nonsingular FRW cosmology and nonlinear electrodynamics}, Phys. Rev. D {\bf69} 123504 (2004).


\bibitem{Novello2007}M. Novello, E. Goulart, J. M. Salim,  S. E. Perez Bergliaffa, {\it Cosmological effects of nonlinear electrodynamics}, Class. Quantum Grav. {\bf24} 3021 (2007).

\bibitem{Kruglov2016}S. I. Kruglov, {\it Acceleration of universe by nonlinear electromagnetic fields}, Int. J. Mod. Phys. D. {\bf 25} 1640002 (2016).

\bibitem{Sarkar2021} P. Sarkar, P. K. Das, G. C. Samanta, {\it Inflationary cosmology- A new approach using Non-linear electrodynamics}, Physica Scripta {\bf96} (2021) 065305.





\bibitem{Bronnikov2006} \textcolor{black}{ K. A. Bronnikov, J. C. Fabris, {\it Regular phantom black holes}, Phys. Rev. Lett. {\bf96}, 251101 (2006).}
%
\bibitem{Bronnikov2007} \textcolor{black}{ K. A. Bronnikov, V. N. Melnikov, H. Dehnen, {\it Regular black holes and black universes}, Gen. Rel. Grav. {\bf39}, 973 (2007). }
%
\bibitem{Bronnikov2010}\textcolor{black}{K.A. Bronnikov, E.V. Donskoy, {\it Possible black universes in a brane world}, Grav. Cosmol. {\bf16} 42 (2010).} 
%
\bibitem{Bronnikov2012} \textcolor{black}{ S. V. Bolokhov, K. A. Bronnikov, M. V. Skvortsova, {\it Magnetic black universes and wormholes with a phantom scalar}, Class. Quantum Gravity {\bf29}, 245006 (2012).}



\bibitem{morris88}
M. S. Morris, K. S. Thorne, {\it Wormholes in Spacetime and their use for Interstellar Travel: A Tool for Teaching General Relativity}, Am. J. Phys {\bf 56}, 395 (1988).

\bibitem{morris88-2}
M. S. Morris, K. S. Thorne, and U. Yurtsever, {\it Wormholes, Time Machines, and the Weak
Energy Condition}, Phys. Rev. Lett {\bf 61}, 1446 (1988).

\bibitem{WEC} M. Visser, {\it Lorentzian Wormholes: from Einstein to Hawking}, AIP Press, New York, 1995; H. Stephani, D. Kramer, M. MacCallum, C. Honselaers, and E. Herlt, {\it Exact solutions to Einstein’s Field Equations}, Second Edition, Cambridge University Press, Cambridge, U.K., 2003.



\bibitem{Plebanski} J. Pleba\'{n}ski, A. Krasi\'{n}ski, {\it An Introduction to General Relativity and Cosmology}, Cambridge University Press (2006).


\bibitem{Ellis} H. G. Ellis, {\it Ether flow through a drainhole: A particle model in general relativity}, J. Math. Phys {\bf 14}, 104 (1973); K. A. Bronnikov, {\it Scalar-tensor theory and scalar charge}, Acta Phys. Pol. B {\bf 4}, 251 (1973)


\bibitem{Visser}M. Visser, {\it Acoustic black holes: horizons, ergospheres and Hawking radiation}, Class. Quantum Grav. {\bf 15}, 1767 (1998).

\bibitem{Unruh} W. G. Unruh, {\it Experimental Black-Hole Evaporation? } Phys. Rev. Lett. {\bf 46}, 1351 (1981).

\bibitem{BHs_CABH} P. Ca\~nate, J. Sultana, D. Kazanas, {\it Gravitational analog of the canonical acoustic black hole in Einstein-scalar-Gauss-Bonnet theory}, Class. Quant. Grav. {\bf38}, 125002 (2021).


\bibitem{4DEGB} D. Glavan, C. Lin, {\it Einstein-Gauss-Bonnet gravity in 4-
dimensional space-time}, Phys. Rev. Lett. {\bf124}, 081301 (2020).


\bibitem{Simpson_2} A. Simpson, P. Mart\'{i}n-Moruno and M. Visser, {\it Vaidya spacetimes, black-bounces,
and traversable wormholes}, Class. Quant. Grav. {\bf36} (2019) 145007. 

\bibitem{Simpson_3} F.S.N. Lobo, A. Simpson and M. Visser, {\it Dynamic thin-shell black-bounce traversable
wormholes}, Phys. Rev. D {\bf 101} (2020) 124035. 


\bibitem{Pe_Fe} P. Ca\~nate, F. H. Maldonado-Villamizar, {\it Novel traversable wormhole in General Relativity and Einstein-Scalar-Gauss-Bonnet theory supported by nonlinear electrodynamics}, arXiv:2202.12463.

\bibitem{Bronnikov_BBs} K. A. Bronnikov, R. K. Walia, {\it Field sources for Simpson-Visser space-times}, Phys. Rev. D {\bf105}, 044039 (2022). 


\end{thebibliography}
\end{document}